\documentclass[
  aps,prl,twocolumn,showpacs,amsmath,amssymb,superscriptaddress
]{revtex4-1}

\makeatletter
\newcommand{\whencolumns}[2]{\preprintsty@sw{#1}{#2}}
\makeatother

\usepackage[english]{babel}
\usepackage{amsmath,amssymb,amsfonts} 	
\usepackage{graphicx}
\usepackage{array,multirow}
\usepackage[utf8]{inputenc}
\usepackage{bm}
\usepackage{xcolor}
\usepackage[normalem]{ulem}
\usepackage{siunitx}
\usepackage{hyperref}
\hypersetup{colorlinks=true,linkcolor=blue,citecolor=blue,urlcolor=blue}
\usepackage{url}
\usepackage{textcomp}
\usepackage{lineno,hyperref}
\usepackage{float}
\usepackage{textcomp} 
\usepackage{rotating}
\usepackage[figurename=FIG.]{caption}
\usepackage{subcaption}
\usepackage{color} 
\definecolor{red}{rgb}{1.,0.,0.}
\usepackage{soul}
\usepackage{amsmath}
\usepackage{enumitem}

 \usepackage[
 justification=raggedright]
 {caption}
\usepackage{subcaption}

\makeatletter
\newcommand*{\rom}[1]{\expandafter\@slowromancap\romannumeral #1@}
\makeatother

\newcommand\THEOS{Theory and Simulation of Materials (THEOS), \'Ecole Polytechnique F{\'e}d{\'e}rale de Lausanne, 1015 Lausanne, Switzerland}
\newcommand\MARVEL{National Centre for Computational Design and Discovery of Novel Materials (MARVEL), {\'E}cole Polytechnique F{\'e}d{\'e}rale de Lausanne, 1015 Lausanne, Switzerland}
\newcommand\MARVELL{National Centre for Computational Design and Discovery of Novel Materials (MARVEL), {\'E}cole Polytechnique F{\'e}d{\'e}rale de Lausanne, 1015 Lausanne, Switzerland}

\newcommand\LIEGE{nanomat/Q-mat/CESAM, Universit\'e de Li\`ege, B-4000 Sart Tilman, Li\`ege, Belgium}

\newcommand\UTRECHT{Chemistry Department, Debye Institute for Nanomaterials Science, Condensed Matter and Interfaces, Utrecht University, PO Box 80.000, 3508 TA Utrecht, The Netherlands}
\newcommand{\ETSF} {European Theoretical Spectroscopy Facility (ETSF) www.etsf.eu}

\begin{document}

\title{Phonon-assisted luminescence in defect centers from many-body perturbation theory}

\author{Francesco Libbi}
\email[Corresponding author. ]{francesco.libbi@epfl.ch}
\affiliation{\THEOS}\affiliation{\MARVEL}
\author{Pedro Miguel M. C. de Melo}
\affiliation{\UTRECHT}\affiliation{\LIEGE}\affiliation{\ETSF}
\author{Zeila Zanolli}
\affiliation{\UTRECHT}\affiliation{\ETSF}
\author{Matthieu Jean Verstraete}
\affiliation{\LIEGE}\affiliation{\ETSF}
\author{Nicola Marzari}
\affiliation{\THEOS}\affiliation{\MARVELL}


\begin{abstract}
Phonon-assisted luminescence is a key property of defect centers in semiconductors, and can be measured to perform the readout of the information stored in a quantum bit, or to detect temperature variations. 
The investigation of phonon-assisted luminescence usually employs phenomenological models, such as that of Huang and Rhys, with restrictive assumptions that can fail to be predictive. 
In this work, we predict luminescence and study exciton-phonon couplings within a rigorous many-body perturbation theory framework,
an analysis that has never been performed for defect centers.
In particular, we study the optical emission of the negatively-charged boron vacancy in 2D hexagonal boron nitride, which currently stands out among defect centers in 2D materials thanks to its promise for applications in quantum information and quantum sensing. We show that phonons are responsible for the observed luminescence, which otherwise would be dark due to symmetry. We also show that the symmetry breaking induced by the static Jahn-Teller effect is not able to describe the presence of the experimentally observed peak at 1.5 eV. \\
\end{abstract}

\maketitle

In the last decade defect centers in semiconductors have been increasingly studied for their applications in quantum computing \cite{Childress281}, quantum communication \cite{Pfaff532,PhysRevLett.96.070504} and quantum sensing \cite{PhysRevLett.103.220802, PhysRevLett.106.080802,PhysRevLett.112.097603,PhysRevLett.112.047601,Neumann2013}. Quantum information is usually stored in a paramagnetic ground state, which is manipulated through electromagnetic radiation, often in the range of microwaves. Optical properties of the defect centers play a major role in quantum information and quantum sensing: spin-dependent luminescence can be used to perform the readout \cite{Toyli8417,PhysRevB.81.035205} of defect centers which are responsive to optically-detected magnetic resonance (ODMR), allowing these systems to be used as practical qubits. Furthermore, the sharp temperature dependence of the peaks of the photoluminescence (PL) spectrum can be used to design sensitive thermometers \cite{Chen2020}. \\
In the last years, a great deal of attention has been focused on
defect centers in 2D hexagonal boron nitride \cite{Kianinia2020,Mendelson2020,Bourrellier2016,PhysRevB.78.155204}, where a wide band gap, together with low spin-orbit coupling and reduced dimensionality, may overcome the limitations of existing 3D defect centers in terms of coherence \cite{Ye2019}, resolution and interfaceability with cavities and resonators \cite{Gottscholl2020}. In particular, the negatively charged boron vacancy ($\mathrm{V_B^-}$) in 2D hBN has been predicted to be responsive to ODMR (and thus usable for quantum computing), and to emit in the infrared when stimulated with continuous green laser light \cite{Gottscholl2020}.\\
Due to the coupling of excitons with vibronic modes of the atoms around a defect, the photoluminescence of defect centers in semiconductors is often phonon-assisted. This is usually studied using the Huang-Rhys model \cite{HR,Alkauskas_2014,Ivady2020,2021,PhysRev.140.A601,doi:10.1063/1.1700283}, whose parameters are determined through constrained-DFT or quantum-chemistry calculations, as reviewed in Ref. \cite{PhysRevB.102.144105}. The Huang-Rhys model, however, is based on restrictive hypotheses (summarised in the supplementary information (SI) \cite{SI} ), such as the assumption that the optical dipole moments are not affected by phonons, which limit its predictive power. Furthermore, it does not lead to an interpretation of the phonon-assisted luminescence in terms of coupling between excitons and phonons, which is essential for the understanding of this phenomenon.\\
The optical properties of \emph{pristine} hBN have been studied using many-body perturbation theory (MBPT) \cite{PhysRevB.97.075121,PhysRevB.98.125206}, and its phonon-assisted emission spectrum has been calculated both for the bulk and the monolayer \cite{PhysRevB.99.081109, PhysRevLett.122.187401,PhysRevLett.122.067401}. However, due to the complexity and computational cost of these analyses, the luminescence of defect centers has never been studied with such advanced methods.\\
%
%
%
%
In the following, we investigate the optical properties of the negatively charged boron vacancy in 2D hBN using first-principles many-body perturbation theory,
and compare with the experimental results of Ref. \cite{PhysRevB.102.144105}. This predictive approach for the luminescence of defect centers gives a universal tool to include the non trivial couplings beyond the Huang-Rhys theory for PL lineshapes, and allows to reach an unprecedented microscopic understanding of the emission mechanism in these systems. In particular, we study the coupling between excitons and phonons leading to the phonon-assisted luminescence, and identify the phonon modes which interact most strongly with excitons, thus leading to peaks in the sideband of the emission spectrum. We show that the phonon-independent emission is dark due to symmetry reasons, and prove that the emission observed experimentally is due to the symmetry-breaking caused by phonons. Conversely, we show that the symmetry-breaking induced by the static Jahn-Teller effect is not able to activate  photoluminescence. The remarkable agreement between theoretical and experimental PL, together with the lack of ad-hoc parameters underscore the predictive accuracy of such approach. \\
\noindent
In a photoluminescence measurement, the system is excited with a laser beam, leading to the formation of excitons. These undergo a series of scattering processes, mainly with phonons and other excitons, as a consequence of which electrons and holes relax respectively towards the bottom of the conduction band and the top of the valence band, before the radiative/non-radiative recombination takes place, with the emission of photons and/or phonons \cite{PhysRevB.92.205304,PhysRevB.93.155102}.
If the excited state dynamics continues for long enough, it is reasonable to assume that both electron and holes thermalise, with pseudo-equilibrium occupations equal to 
\begin{equation}\label{elec}
    f_{n\mathbf{k}} = \frac{1}{e^\frac{\varepsilon_{n\mathbf{k}}-\mu_{el}}{k_BT}+1}\ ,\quad
    \bar{f}_{n\mathbf{k}} = \frac{1}{e^{-\frac{\varepsilon_{n\mathbf{k}}-\mu_{ho}}{k_BT}}+1}
\end{equation}
for electrons and holes respectively. Here $\varepsilon_{n\mathbf{k}}$ is the quasiparticle energy of the state $\{n\mathbf{k}\}$ and $\mu_{el}$ ($\mu_{ho}$) the chemical potential for electrons (holes). The values for $\mu_{el}$ and $\mu_{ho}$, as well as the parameters used in the simulations, are reported in the SI.\\
To proceed, we first determine the absorption spectrum by solving the equilibrium Bethe-Saltpeter equation (BSE) for the two-particle correlation function, as reported in the SI. Then, we calculate the photoluminescent emission by solving the non-equilibrium BSE for the electron-electron correlation function, following the approach of Ref. \cite{PhysRevB.99.081109}. The coupling with phonons as a function of laser frequency $\omega$ and temperature T is included in the simulations via:
\begin{equation}\label{eq_tot_pl}
\begin{aligned}
    I_{ph}(\omega,T) \propto \sum_{\lambda,\nu} \frac{\partial^2 |\Pi_{\lambda}|^2}{\partial x_{\nu}^2} f_{\lambda}^{<}  \Bigl[\delta(\omega -E_{\lambda} -\omega_{\nu}) \frac{n_B(\omega_{\nu},T)}{2\omega_{\nu}}+\\\delta(\omega -E_{\lambda} + \omega_{\nu}) \frac{1 + n_B(\omega_{\nu},T)}{2\omega_{\nu}}\Bigr]\ ,\qquad\qquad
\end{aligned}
\end{equation}
with exciton mode $\lambda$ of energy $E_{\lambda}$ changing with respect to the displacement $x_{\nu}$ induced by a phonon $\nu$ of frequency $\omega_{\nu}$. The second derivative of the dipole moment $\Pi_{\lambda}$ gauges the intensity of the coupling between the exciton and the phonon. Here, $f_{\lambda}^{<}$ is the non-equilibrium exciton occupation function, which is non-vanishing only if the excitons are composed by transitions between bands occupied by excited electrons and holes, and the two Dirac $\delta$ in the square bracket correspond to the cases where an exciton recombines with the creation $\Bigl(\delta(\omega -E_{\lambda} +\omega_{\nu})\Bigr)$ or annihilation $\Bigl(\delta(\omega -E_{\lambda} -\omega_{\nu})\Bigr)$  of a phonon; $n_{B}$ is the Bose-Einstein occupation function for phonon $\nu$ at temperature T.
\begin{figure}[h]
    \centering
            \includegraphics[width=8.5cm]{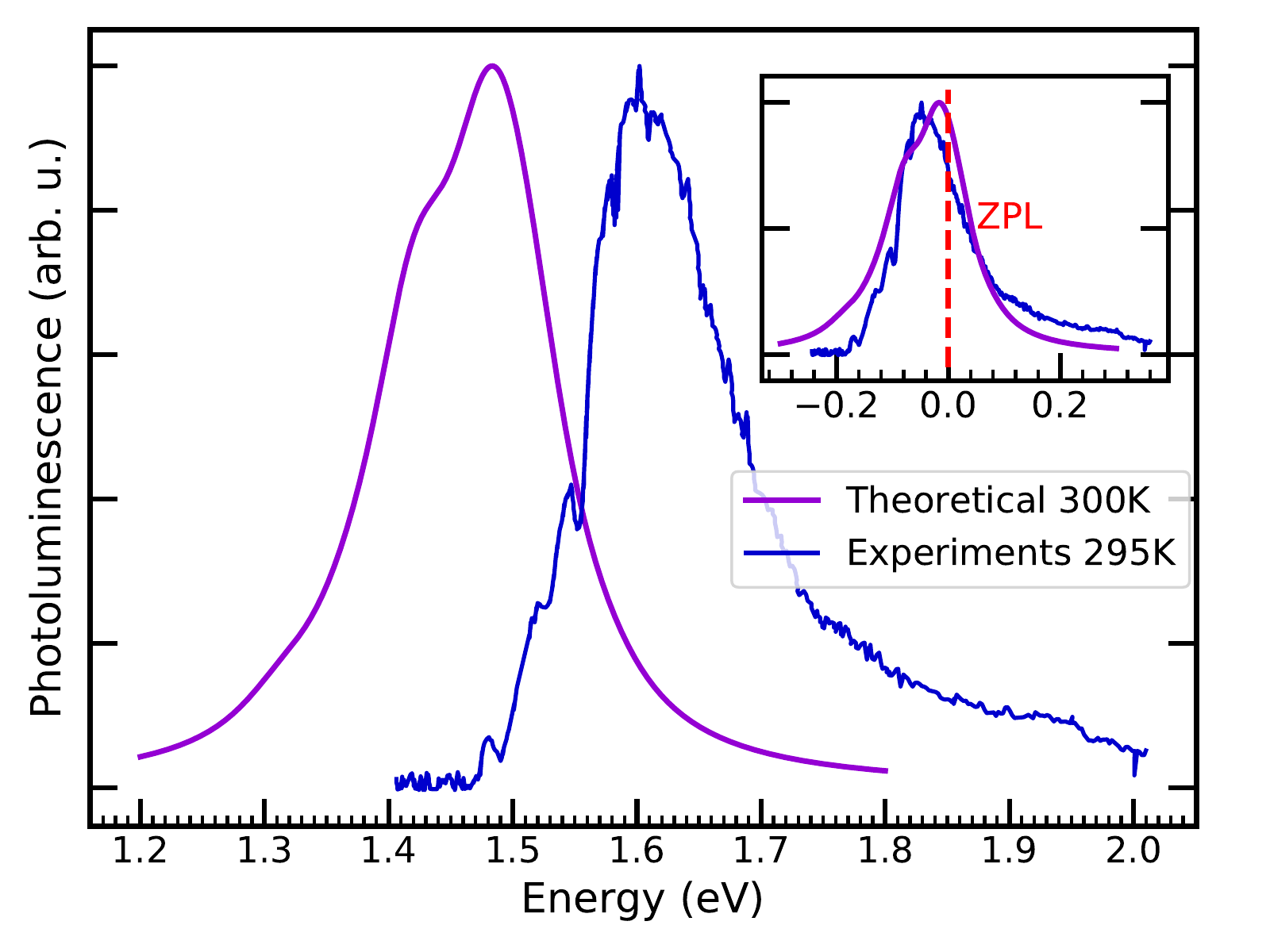}
    \caption{Theoretically predicted phonon-assisted luminescence (Eq. \ref{eq_tot_pl}), normalized and compared to experiments \cite{PhysRevB.102.144105}, at T=300K. The insets shows the theoretical and experimental curves after aligning the ZPLs.}
    \label{fig:tot_pl}
\end{figure}
Eq. \ref{eq_tot_pl} represents a first order effect, the zero-order being the phonon-independent photoluminescence expression:
\begin{equation}\label{pl0}
I_{0} = \sum_{\lambda} |\Pi_{\lambda}^2|f_{\lambda}^{<} 
    \delta(\omega - E_{\lambda} )\ .
\end{equation}
However, as it will be shown later, the phonon-independent luminescence of this system is extremely weak; therefore, it can be neglected and only the first-order term due to phonons survives.
Fig. \ref{fig:tot_pl} compares the theoretically predicted photoluminescence and the experimental one at 300K, showing very good agreement both for the position of the peak and the shape of the spectrum; the inset of Fig. \ref{fig:tot_pl} shows the theoretical and experimental spectra after aligning the zero phonon lines (ZPLs). The experimental ZPL, 1.64 eV at $\mathrm{T=295K}$, is taken from the supplementary information of Ref. \cite{PhysRevB.102.144105}. The theoretical ZPL is assumed to coincide with the energy of the exciton contributing the most to the photoluminescence spectrum (this choice is discussed in detail later). 
In order to identify the contribution of each exciton to the photoluminescence spectrum it is useful to plot a simplified version of Eq. \ref{eq_tot_pl}:
\begin{equation}\label{backb}
    B(\omega) = \sum_{\lambda,\nu} \frac{\partial^2 |\Pi_{\lambda}|^2}{\partial x_{\nu}^2} f_{\lambda}^{<} 
    \delta(\omega - E_{\lambda} ) \ ,
\end{equation}
which is obtained by replacing the Dirac $\delta$ corresponding to the phonon absorption and emission with a single Dirac $\delta$ centered at the exciton energy. This simplified expression has the merit of highlighting the \lq\lq excitonic backbone\rq\rq\ which stands behind the spectra of  Fig. \ref{fig:tot_pl}.
\begin{figure}[h]
    \centering
    \includegraphics[width=8.5cm]{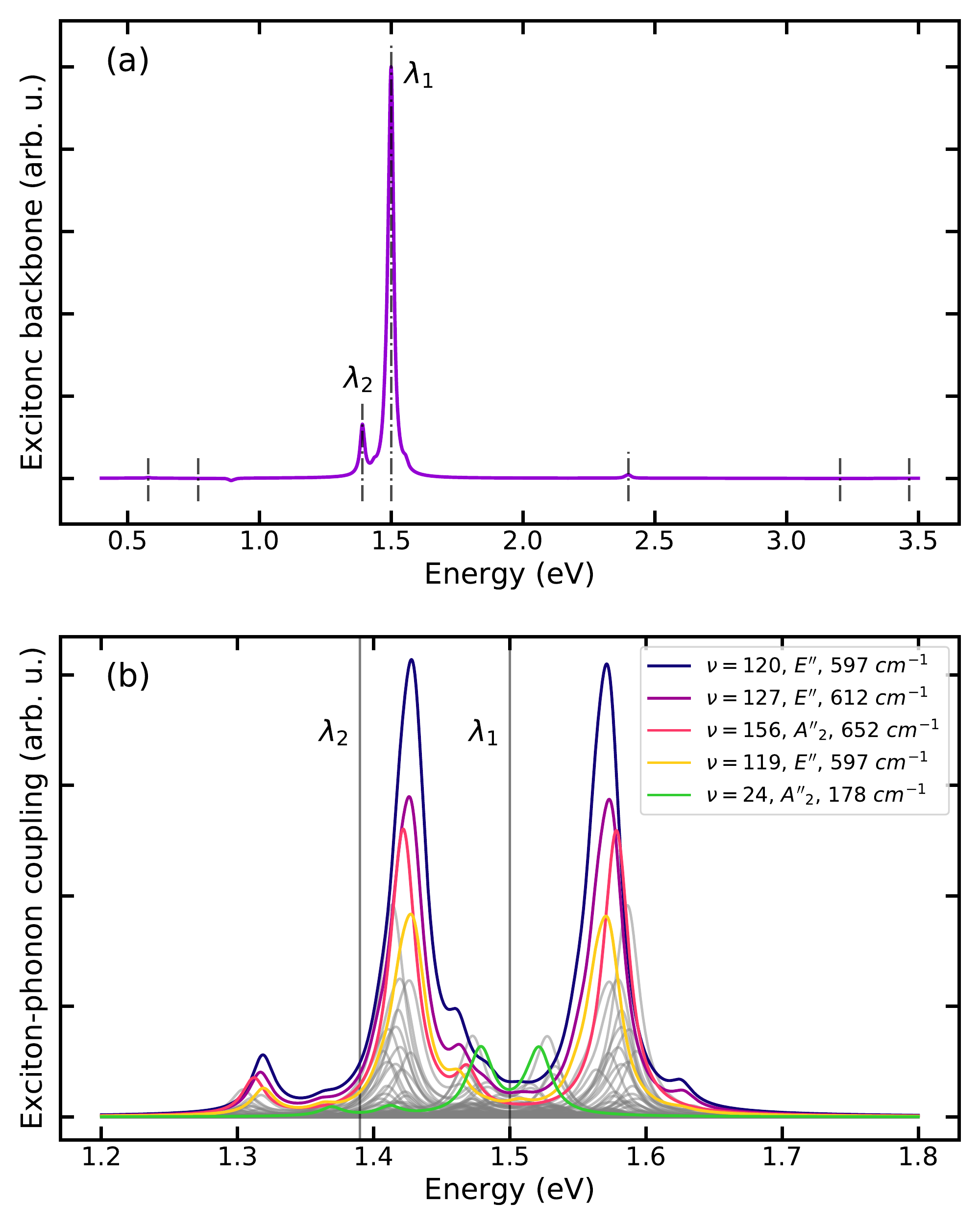}
    \caption{Panel (a) shows the exciton backbone, as defined in Eq. \ref{backb} and normalized. It is clearly dominated by a main excitonic peak at 1.50 eV ($\lambda_1$), which is assumed to coincide with the ZPL of the system, while a smaller detached peak appears at 1.39 eV ($\lambda_2$). Panel (b) represents instead the exciton phonon coupling function, as defined in Eq. \ref{ph-exc} . The phonon modes which couple most strongly with excitons are enlightened with coloured lines and reported in the legend in decreasing order of contribution, while the other phonon modes are represented with grey lines. Mode $\mathrm{\nu=24}$ is not among those coupling most strongly with excitons; however its contribution to the PL will become one of the highest when including the Bose-Einstein terms, as shown in the SI.}
    \label{fig:backbone_ph-exc}
\end{figure}
As shown in Fig. \ref{fig:backbone_ph-exc}\textcolor{blue}{a} this backbone is dominated by a main peak at 1.5 eV ($\lambda_1$) and a smaller peak  at 1.39 eV ($\lambda_2$). The peak at 1.5 eV is distributed into multiple new ones by phonons, whose sum leads to the main peaks of the curve reported in Fig. \ref{fig:tot_pl}. The peak at 1.39 eV in the exciton backbone is instead responsible for the shoulder that appears in the spectra of Fig. \ref{fig:tot_pl} towards 1.3 eV. 
The experimental spectrum is asymmetric with a shoulder at low energy which may correspond to the 1.3 eV peak, though the intensity ratio is different.
%
If the phonons had vanishing frequency, almost all the emission would take place at 1.5 eV, which is the energy of the bare exciton contributing most to the photoluminescence spectrum. Taking this as a definition of the \lq\lq zero phonon line\rq\rq\, then 1.5 eV can be used for a direct comparison with the ZPL of the Huang-Rhys model. The error of 0.14 eV between the experimental (1.64 eV) and theoretical (1.5 eV) ZPLs witness a very good agreement between simulations and experiments.\\
In order to understand which phonon modes couple most strongly with excitons, we plot a simplified version of Eq. \ref{eq_tot_pl}, defining the \lq\lq exciton-phonon coupling\rq\rq\ function:
\begin{equation}\label{ph-exc}
    C(\omega) = \sum_{\lambda,\nu} \frac{\partial^2 |\Pi_{\lambda}|^2}{\partial x_{\nu}^2} f_{\lambda}^{<} \Bigl[
    \delta(\omega - E_{\lambda} -\omega_{\nu})  +
    \delta(\omega - E_{\lambda} +\omega_{\nu})  
    \Bigr] \ 
\end{equation}
obtained from Eq. \ref{eq_tot_pl} by removing the Bose-Einstein distributions, which otherwise would affect the amplitudes of the peaks and would depend on temperature.
We can see from Fig. \ref{fig:backbone_ph-exc}\textcolor{blue}{b} that the mode that couples most with excitons is $\mathrm{\nu=120}$, of frequency 597 $\mathrm{cm^{-1}}$ and symmetry $E''$, which is characterised by an out-of-plane oscillation of the first and second nearest neighbours around the vacancy (as shown in the SI). The other phonon modes which dominate the exciton-phonon coupling function have similar frequencies to mode $\mathrm{\nu=120}$ and either $E''$ or $A_2''$ symmetry (mode $\nu=24$, charachterized by in-phase out-of-plane oscillation of the nitrogen atoms around the vacancy). This ranking is overturned by the Bose-Einstein occupations, especially at high temperatures. As discussed in detail in the SI, the effects of the terms $\frac{n_B(\omega,T)+1}{2\omega}$ and $\frac{n_B(\omega,T)}{2\omega}$ are (i) to increase the contribution to PL of the low energy phonons at high temperatures, (ii) to suppress the contribution of the phonon annihilation term at low temperatures. 

%
%
\begin{figure}[b]
    \centering
    \includegraphics[width=8.5cm]{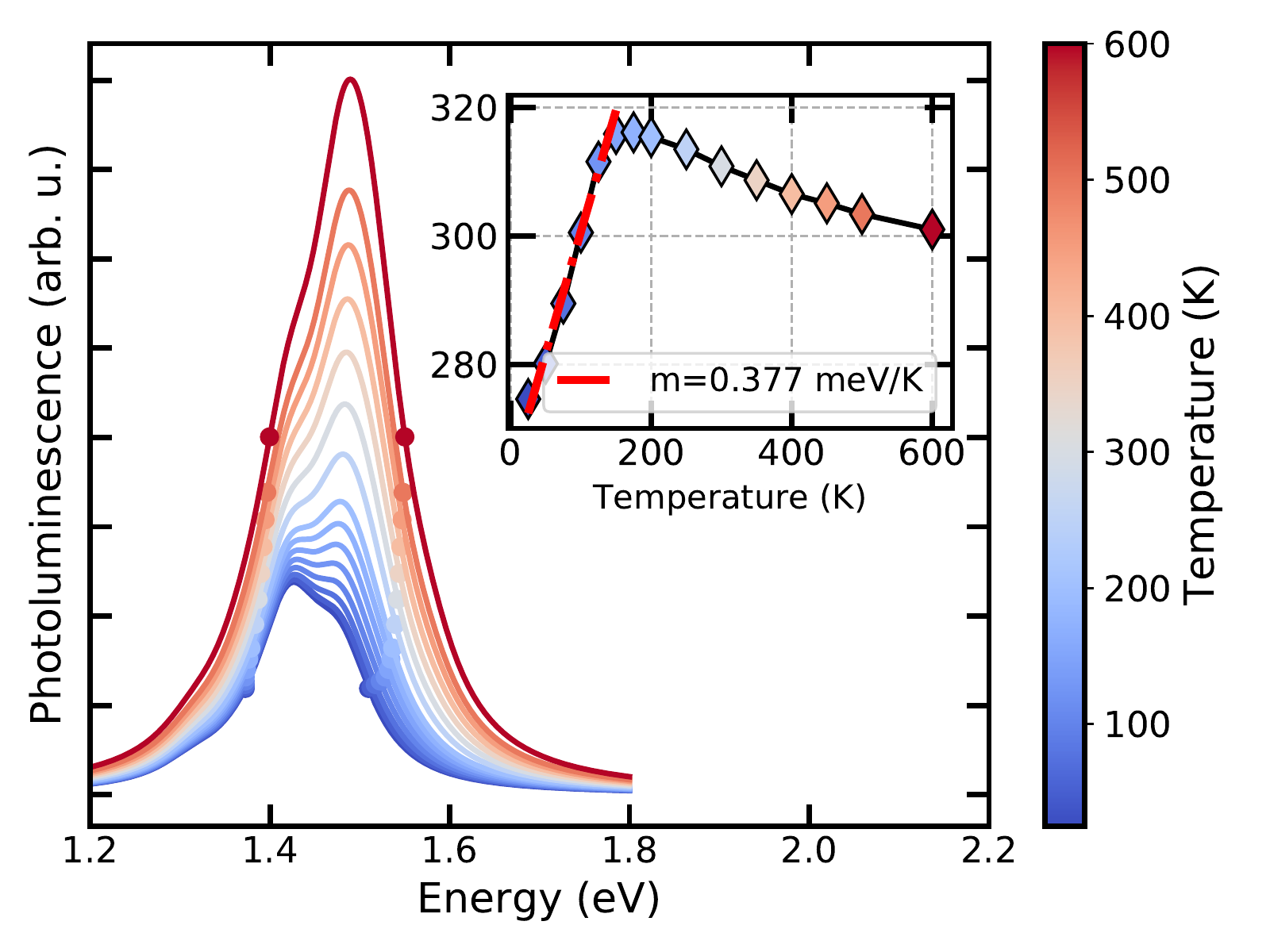}
    \caption{Temperature dependence of the phonon-assisted luminescence spectrum. The pair of dots which are plotted for each spectrum show the points where it has a value which is half its maximum (full width at half maximum - FWHM). The inset shows the behaviour of this FWHM (in meV) as a function of temperature. Note that the increase of the curves along the vertical axis is not quantitative, as Eq. \ref{eq_tot_pl} does not fix the amplitude of the PL spectrum: only the shape and width are relevant.}
    \label{fig:temperature}
\end{figure}
\begin{figure}[t!]
    \centering
    \includegraphics[width=8.5cm]{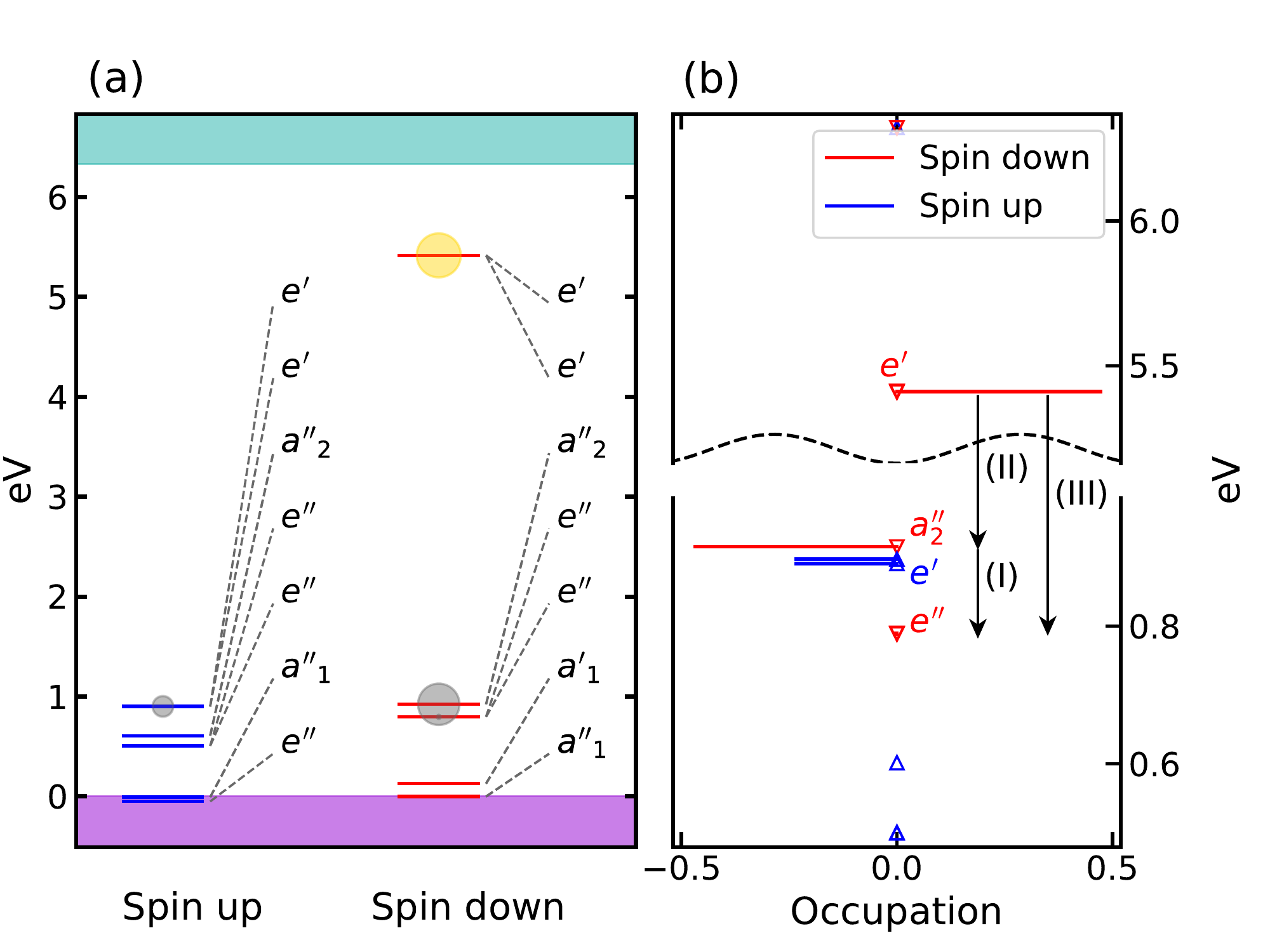}
    \caption{Panel (a) shows the defect levels inside the band-gap for $\mathrm{V_B^-}$ center in 2D hBN, determined through $\mathrm{G_0W_0}$ calculations. The blue and red lines represent spin up and spin down levels respectively. The yellow and grey circles represent pictorially the non-equilibrium occupations of the levels. Panel (b) shows the non-equilibrium occupations of the defect levels as a function of their non-renormalized energies. The black arrows mark the only possible recombinations which can give rise to photoluminescence.}
    \label{defect_occupations}
\end{figure}
\noindent
Chen et al. \cite{Chen2020} suggest to use defects in 2D hBN as sensitive local thermometers, by exploiting the temperature dependence of the full-width at half-maximum (FWHM) of the PL. We calculate the FWHM for a set of temperatures ranging from 0 to 600 K (Fig. \ref{fig:temperature}) and observe an initial very steep increase, due to the combined effects of the increasing contribution of the phonon annihilation term and the increasing importance of lower energy modes.
Eventually the low frequency (178 $\mathrm{cm^{-1}}$) phonon mode $\mathrm{\nu=24}$ of symmetry $A''_2$ starts to dominate the PL spectrum. 
Lower $\omega$ lead to smaller energy splittings, and beyond 200K our exciton-phonon coupling model predicts a smooth decrease in the FWHM, but we emphasize that we are neglecting anharmonic phonon lifetimes as well as multi-phonon effects beyond the perturbation theory used here. These would add to the full spectral peak width, but are of very different nature.
%
A linear fit of the FWHM from 0 to 200K yields an coefficient of 0.343 meV/K, which shows very high sensitivity to temperature variations and suggests that this defect can indeed be used as a nano-scale thermometer.

\begin{figure}[t]
    \centering
    \includegraphics[width=8.5cm]{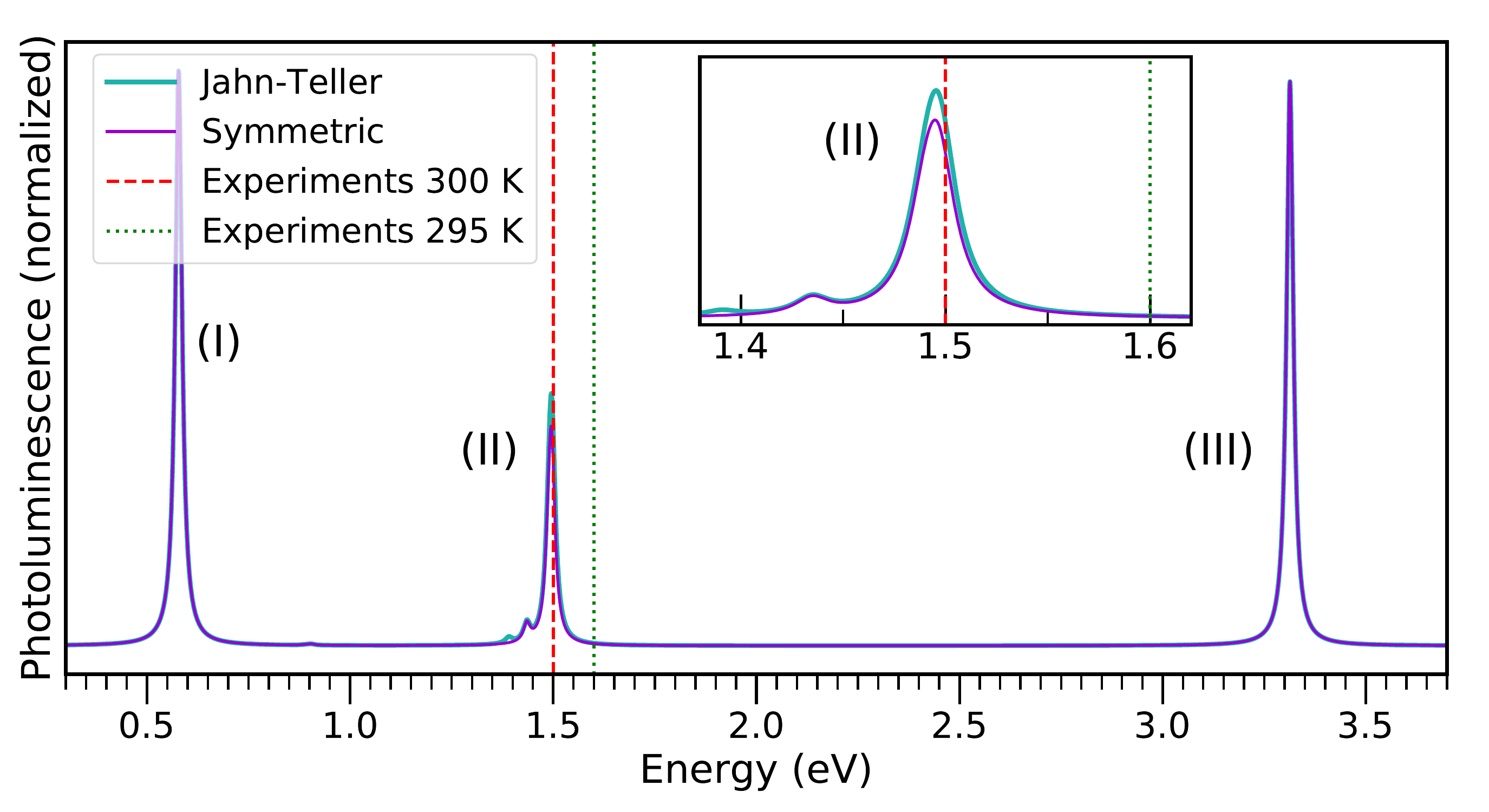}
    \caption{The phonon-independent photoluminescence spectrum of the original symmetric system (purple solid line) is compared to that of the Jahn-Teller distorted system (light blue solid line). The roman numbers indicate which of the recombinations reported in Fig. \ref{defect_occupations}\textcolor{blue}{b} is associated to the peak. The activation of the transition $a_1\rightarrow b_2$ caused by the symmetry breaking due to the Jahn-Teller effect leads to an enhancement of the peak at 1.5 eV, which, however, is still weaker than the other two peaks and therefore not observable by experiments.}
    \label{fig:jt_spectrum}
\end{figure}

%
%
In the final part of this work we argue that mechanisms other than phonon dressing can not explain the observed photoluminescence. We calculate first the phonon-independent luminescence from Eq. \ref{pl0}, and interpret it by studying the defect levels involved, and their non-equilibrium occupations.
The negatively charged boron vacancy generates many donor states close to the valence band, but just one doubly degenerate spin down acceptor close to the conduction band (Fig. \ref{defect_occupations}\textcolor{blue}{a}). The presence of this acceptor opens the spin-down transition channels inside the band-gap, which will be responsible for the observed photoluminescence, as discussed later.
In Fig. \ref{defect_occupations}\textcolor{blue}{b} we show the non-equilibrium occupation of the defect levels (Eq. \ref{elec}) as a function of their quasiparticle energy at 300 K. 
Only the energy levels closest to the middle of the band gap will show significant populations (with a thermal width of 26 meV).
The spin down degenerate levels $e'$ are occupied on average by one excited electron (0.5 each), while the spin down $a_2''$ level by close to 0.5 holes. The spin down level $e''$ has a very small but not negligible population (0.006 holes), while all the other spin down levels are empty.\\
The phonon-independent photoluminescence spectrum reported in Fig. \ref{fig:jt_spectrum} is made up of three main peaks: the first peak at 0.6 eV (I in Fig. \ref{fig:jt_spectrum}) is mainly associated to the spin down $a_2''\rightarrow e''$ transition, which has a non vanishing in-plane dipole. Due to the very small hole population of the level $e''$, this peak is expected to be extremely weak, and is not observed in experiments;
for the same reason the third peak at 3.3 eV (III in Fig. \ref{fig:jt_spectrum}) is very weak and not observed, despite the finite out-of-plane dipole moment of the transition. It is associated to the transition between the spin down levels $e'$ and $e''$.
Last, the second peak at 1.5 eV (II in Fig. \ref{fig:jt_spectrum}) is associated to the transition from the spin down level $e'$ to $a_2''$,
which is the only one where both the levels involved have a non-negligible population in terms of electrons/holes. However, this transition is prohibited by symmetry, therefore the corresponding peak is even weaker than the others.  It is important to stress that the creation of this exciton from the transition $e' \rightarrow a_2''$ is achieved thanks to non-equilibrium occupations: there is no trace of an exciton at 1.5 eV with the same composition when solving the equilibrium BSE.
So, two of the three peaks are very weak because of population arguments, and are not observed in experiments, while a third peak, which has the same energy of the peak observed experimentally (1.5 eV), is dark due to symmetry reasons. Following this analysis, we can conclude that there must be some symmetry-breaking phenomenon which enables the transition $e'\rightarrow a''_2$ (the only one which is not removed by the lack of electron/hole population) and leads to the peak at 1.5 eV which dominates the photoluminescence spectrum, and makes it observable by experiments. The two main candidate mechanisms to activate the second peak are the Jahn-Teller distortion of the excited state $e'$, and lattice vibrations.\\
We have seen above in Fig. \ref{fig:backbone_ph-exc}\textcolor{blue}{a} that the symmetry reduction associated to lattice vibrations allows the transition $e'\rightarrow a_2''$, thus leading to the peak at 1.5 eV to dominate the exciton backbone spectrum.
We now show that the static Jahn-Teller effect, instead, is not enough to activate the second peak.
The doubly degenerate excited state which is obtained promoting an electron from $a''_2$ to  $e'$ can undergo a Jahn-Teller distortion which reduces the symmetry of the system from $D_{3h}$ to $C_{2v}$, splits the level $e'$ into $a_1$ and $b_1$, and changes the irreducible representation of the level $a''_2$ into $b_2$ (see Fig. 8 in the SI). The transition $a_1 \rightarrow b_2$ has now a non-vanishing out of-plane-dipole, while the transition $b_1 \rightarrow b_2$ is still forbidden. 
We calculate the luminescence for the Jahn-Teller distorted system.
As shown in Fig. \ref{fig:jt_spectrum}, the activation of the transition $a_1\rightarrow b_2$ caused by the symmetry break leads to an enhancement of the peak at 1.5 eV. This enhancement, however, is still not sufficient, as this peak remains weaker than the other two peaks, which are not observable as discussed before.
Therefore, we conclude that the static Jahn-Teller effect is not the symmetry breaking phenomenon which activates the photoluminescence and the peak observed at 1.5 eV. A tell tale signature for this reasoning can be also found in the lack of an apparent ZPL in the experimental spectra. 

In conclusion, we studied the phonon-assisted luminescence of the negatively charged boron vacancy in 2D hBN using MBPT. Our simulations show that the phonon-independent luminescence is extremely weak, even when the static Jahn-Teller effect is included. Instead, phonons of $E''$ and  $A_2''$ symmetry are responsible for the observed luminescence.
At low temperatures, the luminescence is determined by high frequency modes which are strongly coupled with excitons. At $\mathrm{T>200K}$ the PL spectrum becomes dominated by low-frequency modes due to the Bose-Einstein statistics. The temperature dependence of the FWHM shows a very sensitive behaviour of the PL spectrum with the temperature, suggesting that this defect can be used as a nanoscale thermometer, at least in a range of temperatures between 0 and 200 K. The agreement between theory and experiments is very good, and we hope our explanation of the photoluminescence mechanisms will further the technological integration of defect centers for quantum information and sensing.

\begin{acknowledgments}
This project has received funding from the European union's Horizon 2020 research and innovation program under the Marie Skłodowska-Curie grant agreement \textnumero$\ $754354 and was supported by the NCCR MARVEL, funded by the Swiss National Science Foundation. 
MJV and PMMC were funded by the Fonds de la Recherche Scientifique (FRS-FNRS Belgium) through PdR Grant No. T.0103.19 - ALPS.
Simulation time was awarded by PRACE on Marconi at Cineca, Italy (project id. 2016163963) and by PRACE \textit{optospin} on Mare Nostrum at Barcelona Supercomputing center, Spain (project id. 2020225411).
Zeila Zanolli  and Pedro Miguel M. C. de Melo acknowledge financial support by the Netherlands Sector Plan program 2019-2023.
Matthieu Jean Verstraete acknowledges funding from Federation Wallonie Bruxelles through Actions de Recherches Concertées project DREAMS g.a. ARC 21/25-11.
\end{acknowledgments}


%

\end{document}


\title{Supplementary information}

\maketitle

\section*{Calculation parameters}
\noindent
In order to determine the ground state properties of the negatively charged boron vacancy in 2D hBN we perform DFT calculations using the PBE exchange-correlation functional on a 8x8 supercell with the open-source Quantum ESPRESSO distribution \cite{Giannozzi_2009}. The planewave cutoff used is 80 Ry and the Brillouin zone is sampled with a 2x2x1 unshifted grid. The norm-conserving pseudopotential adopted for boron and nitrogen are taken from the PseudoDojo library \cite{VANSETTEN201839}.\\
We correct the Kohn-Sham eigenvalues with a $\mathrm{G_0W_0}$ calculation under the plasmon pole approximation, using the Yambo code \cite{MARINI20091392,Sangalli_2019}. We use 5 Ry and 19 Ry cutoff respectively for the response function and the exchange self energy, and sum 2000 bands both for the response function and the correlation self-energy. We sample the Brillouin zone with a 2x2x1 unshifted grid. We checked the results by performing the same calculations on a 10x10 supercell, sampling the Brillouin zone with a gamma-only grid, summing 3000 bands both for the response functions and the correlation self-energy and using 4 Ry and 15 Ry cutoff respectively for the response function and the exchange self energy.\\
The absorption spectrum is determined by solving the equilibrium Bethe-Saltpeter equation (BSE), using 2 Ry cutoff for the electron-hole attraction part of the kernel, 9 Ry cutoff for the exchange part of the kernel and bands from 190 to 320 (whose energy is corrected using the $\mathrm{G_0W_0}$ approximation). The FFT grid is reduced to 21 Ry. Results were 
The phonon-assisted luminescence calculations converge when considering bands from 251-256, thanks to the effect of the nonequilibrium occupations.\\
The phonons are calculated in $\mathrm{\Gamma}$ with the finite-difference approach, using the code Phonopy \cite{phonopy}.
\clearpage

\section*{Absorption}
\begin{figure}[h]
    \centering
    \includegraphics[scale=0.5]{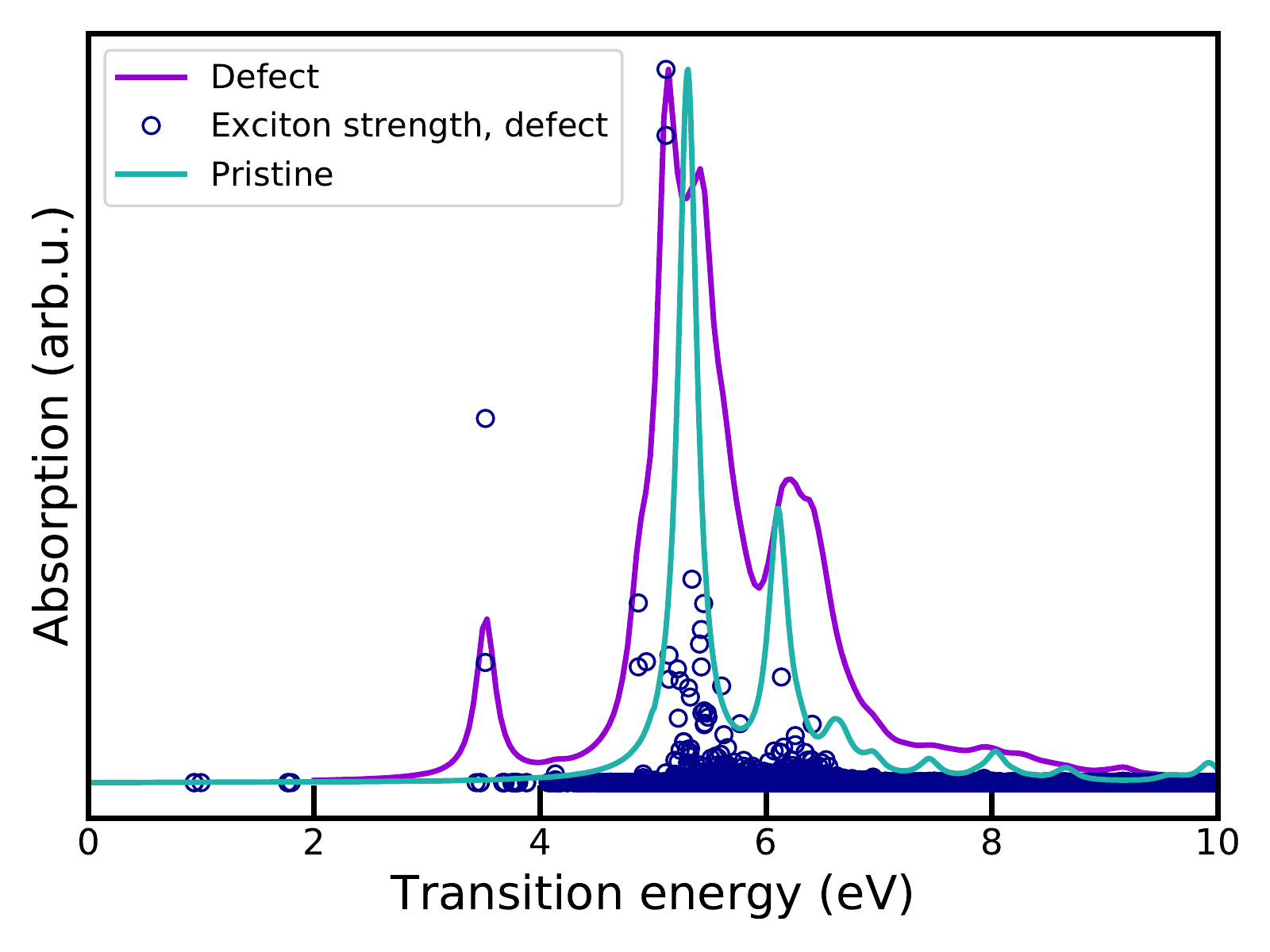}
    \caption{Absorption spectrum for $\mathrm{V_B^-}$ center in 2D hBN (purple line) and for the pristine 2D hBN (light blue line). The blue circles represent the strength of the excitons of the defected system as a function of their energy.}
    \label{abs_spectrum}
\end{figure}
\noindent
The absorption spectrum is shown in Fig. \ref{abs_spectrum}.
Here we perform an analyisis similar to that done in Ref. \cite{PhysRevB.83.144115} for the absorption spectrum of the neutral defects in 2D hBN. 
The strong peak at 5.3 eV which dominates the pristine spectrum, is splitted into two subpeaks at 5.15 eV and 5.40 eV by the resonant coupling of transitions between the extended conduction and valence states with the defect states close to the band edges.\\ 
The transition between the spin down level $a_1'$ and the spin down level $e'$ (see Fig. 4a of the main text), which is associated to a non vanishing in-plane dipole, leads to a bright peak at 3.5 eV. The shoulder at 4.85 eV arises due to the mixing of the transitions from the defect levels to the conduction band and the bulk excitations.\\
The excitons at 0.94 eV and 1.77 eV do not contribute appreciably to the absorption spectrum because are localised on the transitions $e''\rightarrow e'$, which has a non-vanishing out-of-plane dipole and therefore is very weak in an atomically thin material, and $a''_2\rightarrow e'$, which is prohibited by symmetry.

\clearpage
\section*{Effect of nonequilibrium occupation}
\begin{figure}[h]
    \centering
    \includegraphics[scale=0.5]{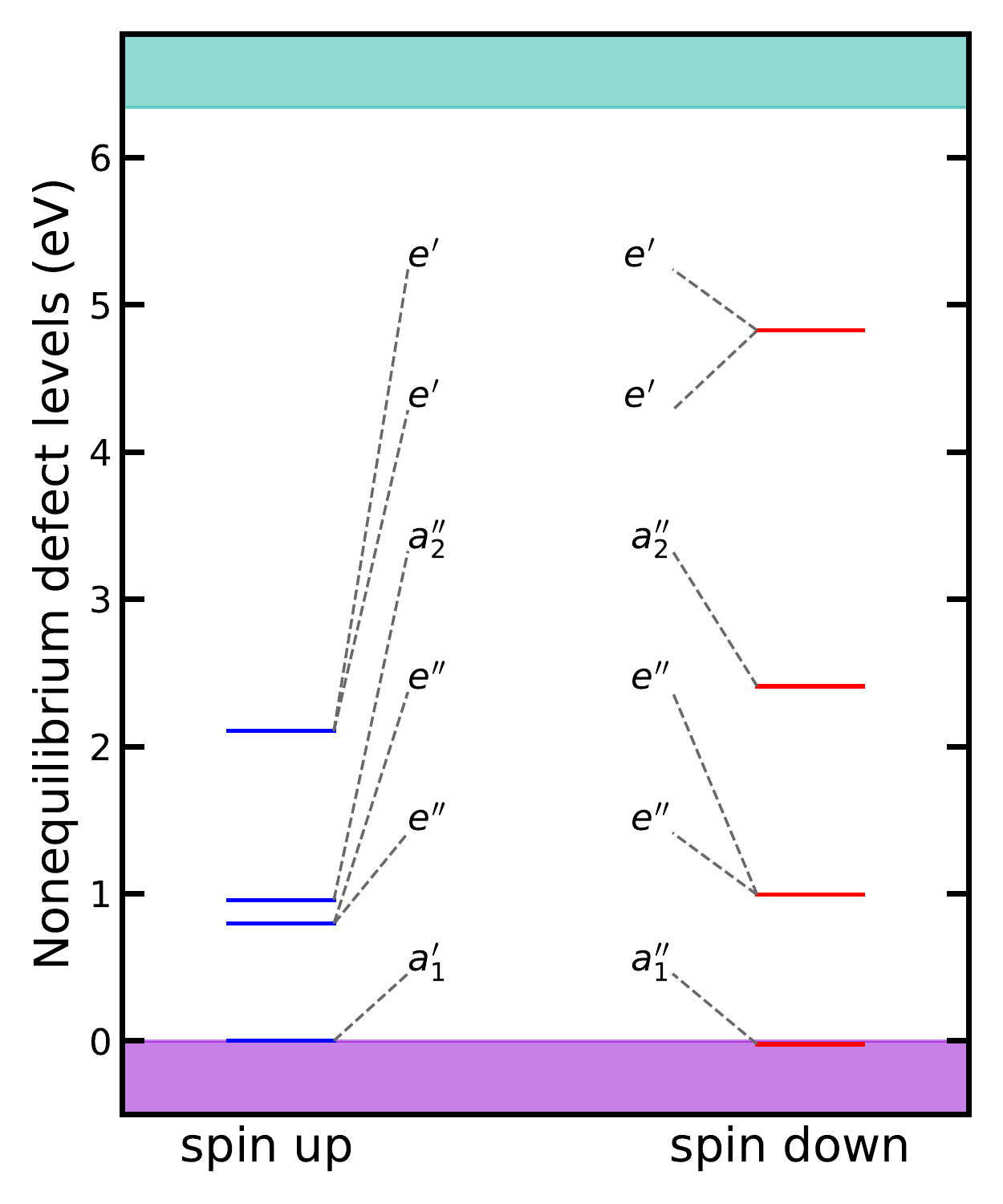}
    \caption{Nonequilibrium energy levels}
    \label{cohsex}
    \caption{Defect energy levels renormalized by the nonequilibrium occupations of Fig. 4a of the main text.}
    \label{fig:occupations}
\end{figure}
\begin{figure}[h]
    \centering
    \includegraphics[scale=0.5]{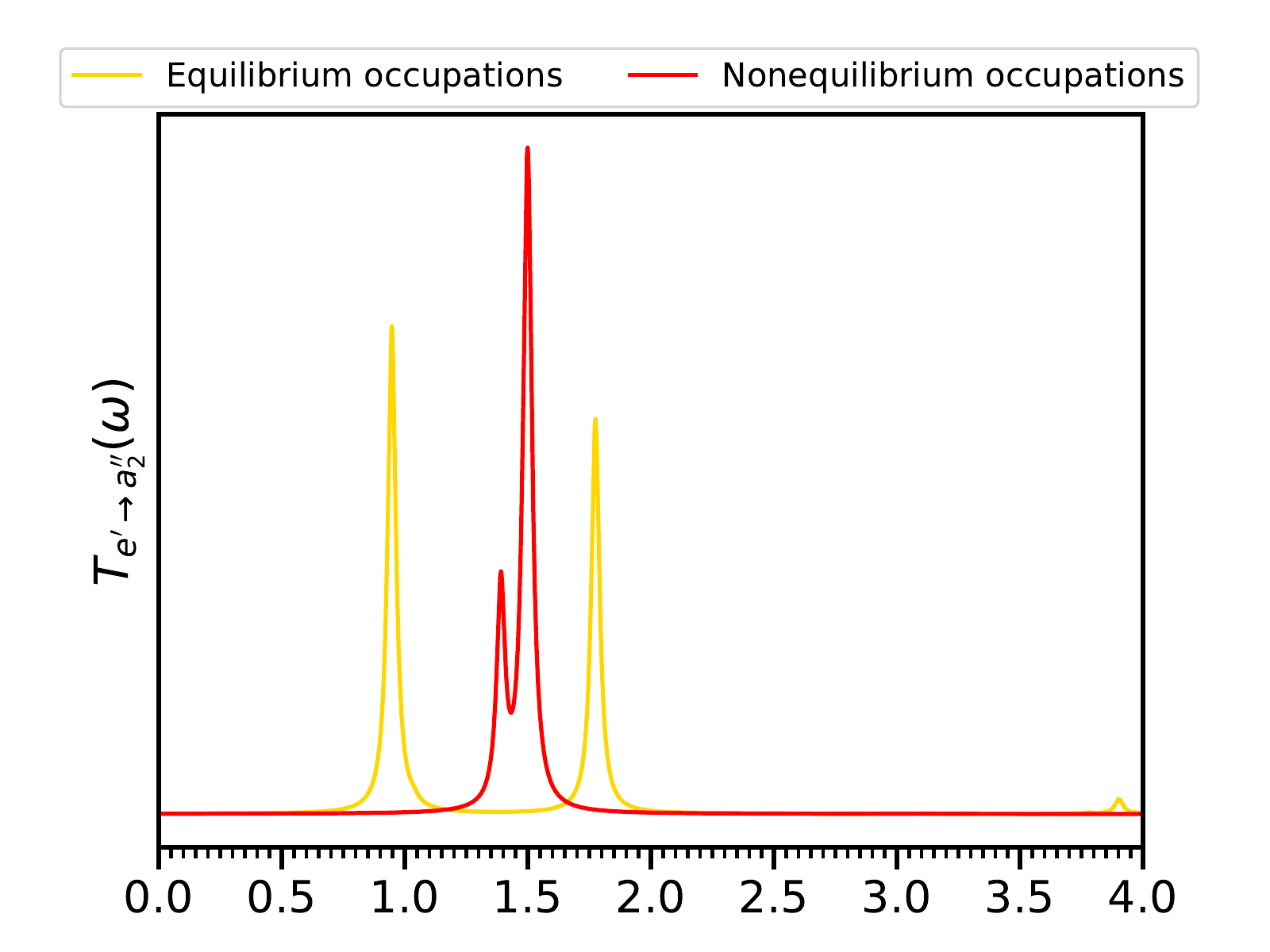}
    \caption{Distribution of the transition $e'\rightarrow a_2''$ among the excitons, both at equilibrium and out of equilibrium.}
    \label{fig:localisation}
\end{figure}
\begin{figure}
    \centering
    \begin{subfigure}{0.48\textwidth}
        \includegraphics[scale=0.5]{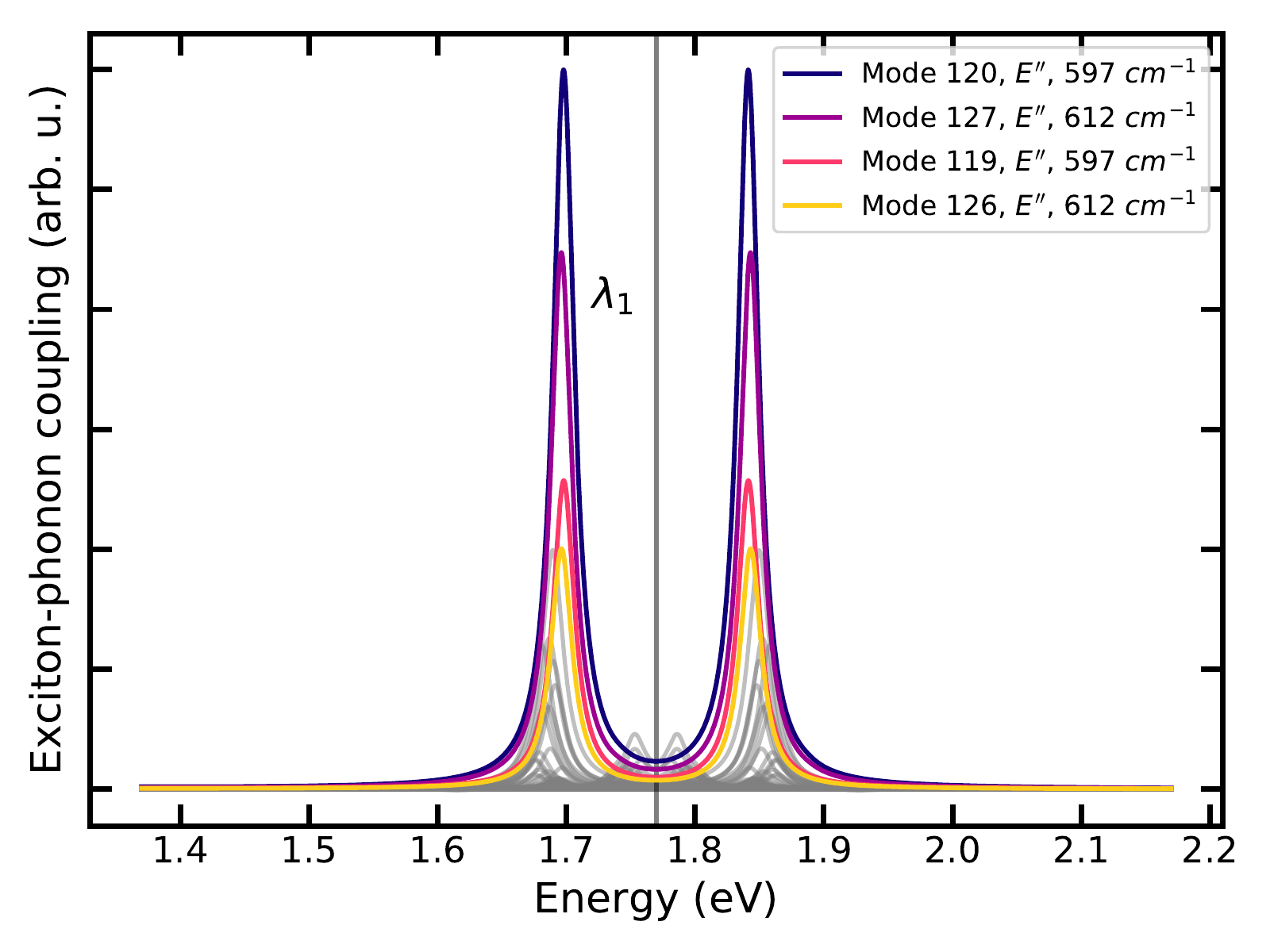}
        \caption{Exciton-phonon coupling}
        \label{ph_exc_eq}
    \end{subfigure}
    \begin{subfigure}{0.48\textwidth}
        \includegraphics[scale = 0.5]{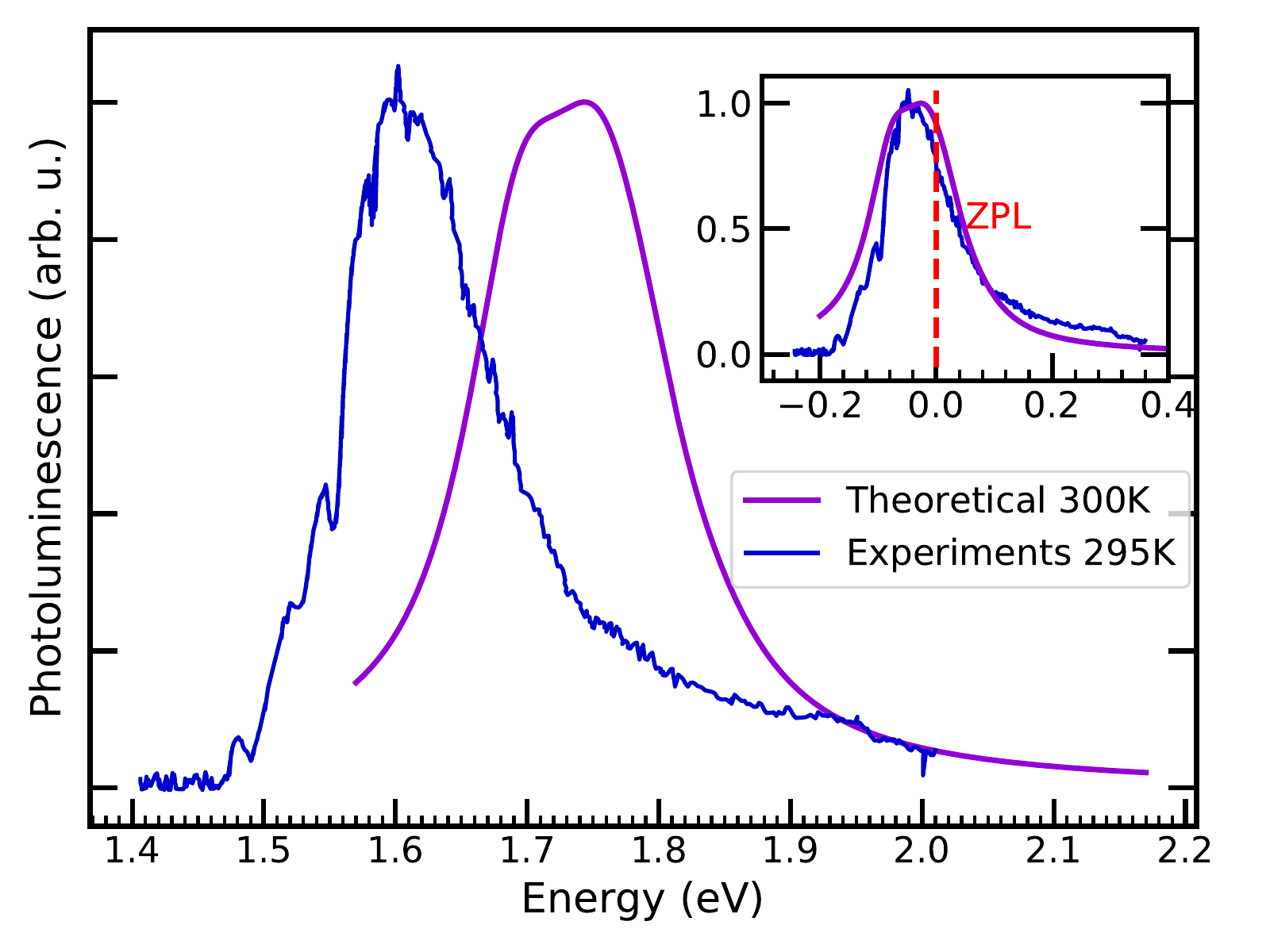}
        \caption{Luminescence spectrum}
        \label{pl_eq}
    \end{subfigure}
    \caption{Panel (a) shows the exciton-phonon coupling function in the case of equilibrium occupations (very low excited charge) for the exciton at 1.77 eV, while panel (b) shows the luminescence spectrum.}
    \label{pl_eq_tot}
\end{figure}

As stated in the main text, if the excited state dynamics continues for long enough, it is reasonable to assume that both electron and holes thermalise.
In the current work, we set the chemical potential for the electrons in such a way that a whole electron is promoted to the excited state manifold, consistently with what is done in works based on quantum chemistry techniques or cDFT \cite{PhysRevB.102.144105}. This corresponds to an excited carrier density of $\mathrm{2.5\times10^{13}}$ electrons per $\mathrm{cm^2}$. The temperature of electrons and holes is assumed to be equal to that of the phonon bath with which they interact. The chemical potential of the holes is tuned in such a way that the number of holes coincides with that of the electrons excited in the conduction band. This assumes the excitations in photoluminescence experiments are neutral. \\
The nonequilibrium occupations induce a renormalization of the quasiparticle energy levels. The renormalized energies are calculated in Yambo as:
\begin{equation}
    \varepsilon_{NEQ} = \varepsilon_{EQ}^{G0W0} + (\varepsilon_{NEQ}^{COHSEX} - \varepsilon_{EQ}^{COHSEX})\, ;
\end{equation}
The first therm in the r.h.s. is the quasiparticle energy determined at the $G_0W_0$ level of theory using the equilibrium occupations (Fig. 4a of the main paper), while the second and the third term represent the quasiparticle energy at the COHSEX level of theory calculated using the nonequilibrium and equilibrium occupations respectively.
As it can be observed when comparing Fig. \ref{cohsex} with Fig. 4a of the main paper, the nonequilibrium occupations lead to a downshift of the spin down level $e'$ which is occupied by excited electrons, and a upshift of the spin levels $e''$ and $a_2''$ which are occupied by holes.\\
As discussed in the main work, the only transition where both levels have non-negligible population in term of electrons and holes is the (spin down) $e'\rightarrow a''$. Therefore it is interesting to see how this transition is distributed among the excitons, both at equilibrium and at nonequilibrium, by plotting
\begin{equation}
    T_{K}(\omega) = \sum_{\lambda} |A_K^{\lambda}|^2\delta(\omega-E_{\lambda})
\end{equation}
for $K\equiv e'\rightarrow a_2''$. As it is possible to observe in Fig. \ref{fig:localisation}, at equilibrium this transition is mainly localised in exciton of energy 0.94 eV and 1.77 eV.  When an integer amount of charge is excited, so that the nonequilibrium occupations differ from the equilibrium ones, the transition $e'\rightarrow a_2''$ localises on two excitons of energy 1.4 eV and 1.5 eV.
It is instructive, finally, to plot the phonon-assisted photoluminescence and the exciton-phonon coupling in the equilibrium occupations case for the exciton at 1.77 eV (see Fig. \ref{pl_eq_tot}).
Interestingly, the modes which couple the most strongly with excitons are the same, leading to a similar luminescence spectrum. The exciton-phonon coupling is now symmetric, due to the absence of a excitonic shoulder at lower energy. 

\clearpage
\section*{Phonon calculations}
As anticipated in the first section, phonons are calculated at the $\mathrm{\Gamma}$ point only using a finite-difference approach. The other points of the Brillouin zone are not considered in the exciton-phonon coupling calculations, as they are characterized by a periodicity which is unphysical for a defect center.
The 8x8 supercell contains 127 atoms, leading to 381 phonon modes, whose frequencies range from 0 to around 1500 $\mathrm{cm^{-1}}$. Due to numerical noise in the calculations, 7 membrane modes have slightly negative frequencies, and are not included in the exciton-phonon coupling calculations.\\
It is interesting to analyze the localization of the phonon modes around the defect centers.
\begin{figure}[h]
    \centering
    \includegraphics[scale=0.5]{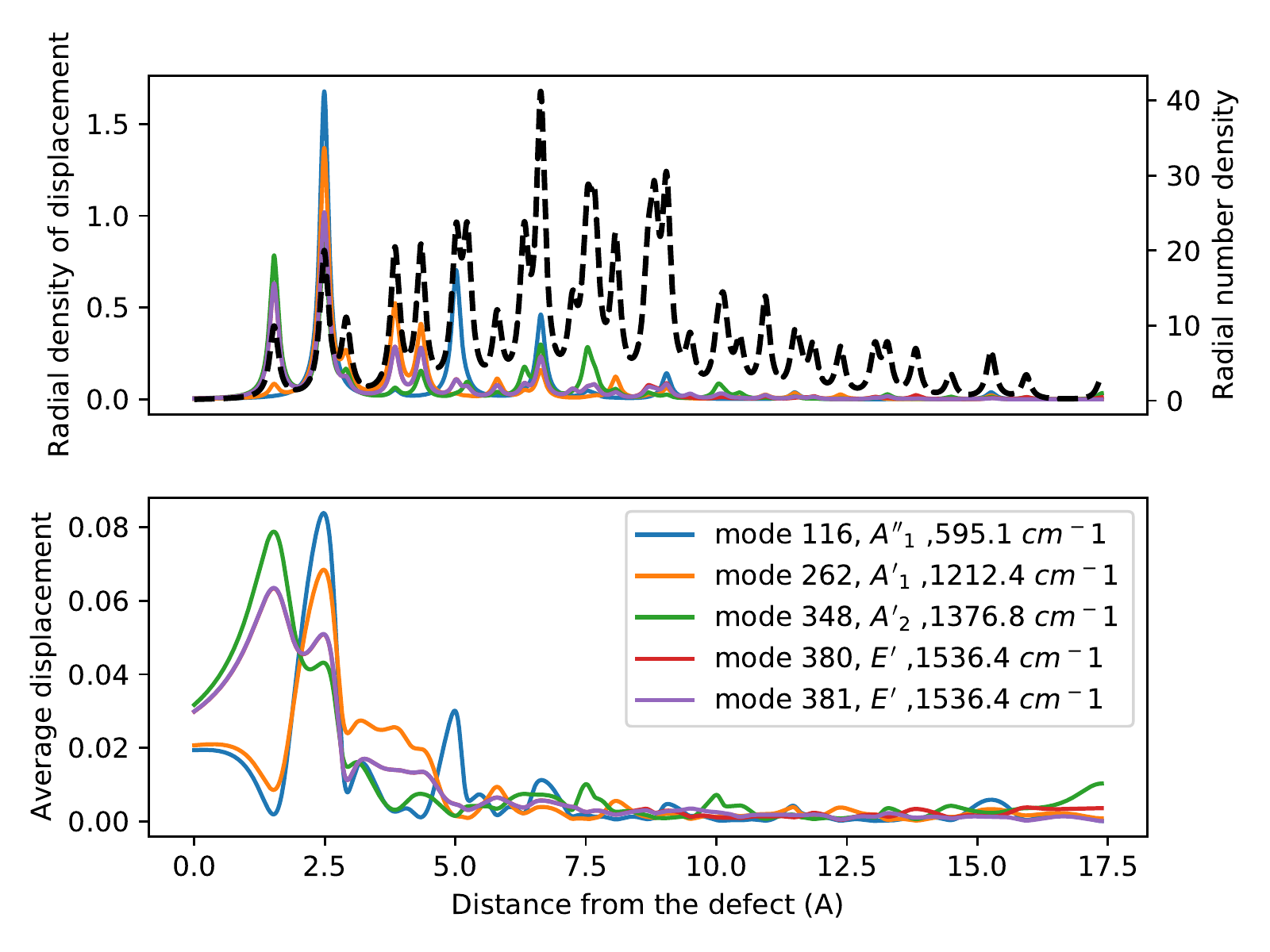}
    \caption{The upper panel shows the radial density of displacement $d_{\nu}(r)$ as defined in Eq. \ref{rdd} for the phonons with the highest localisation factor $f^{\nu}$. The black dashed line indicates the radial number density. The lower panel represents the average displacement as defined in eq. \ref{av_d}.}
    \label{ph-loc}
\end{figure}
With this purpose, we calculate a localization factor defined as the sum of the squared modulus of the phonon eigenvector $\xi_{\nu}$ for the atoms up to the second nearest neighbors around the vacancy:
\begin{equation}
    f^{\nu} = \sum_{i <= 2^{nd} nn} |\xi_{\nu}^i|^2 \ .
\end{equation}
The phonons modes with a localization factor higher than 50$\%$ are (in decreasing order of $f^{\nu}$) $\nu$=116, 262, 348, 380, 381.
Their radial density of displacement
\begin{equation}\label{rdd}
    d_{\nu}(r) = \sum_i |\mathbf{\xi}_{\nu}^{i}|^2 \delta(|\mathbf{r}_i - \mathbf{r}_{vac}|)\ ,
\end{equation}
the radial number density
\begin{equation}\label{nd}
    n(r) = \sum_i \delta(|\mathbf{r}_i - \mathbf{r}_{vac}|)\ ,
\end{equation}
and the average density of displacement
\begin{equation}\label{av_d}
    d_{\nu}^{av}(r) = \frac{d_{\nu}(r)}{n(r)}
\end{equation}
are represented in Fig. \ref{ph-loc} (Here the index i refers to the atomic sites and $\mathbf{r}_{vac}$  is the vacancy position).
It is easy to see that these very local phonon modes do not coincide with those dominating the exciton-phonon coupling spectrum of Fig. 2b of the main text, although the phonon mode $\mathrm{\nu=120}$ as a localization factor of $40.1\%$. Therefore we conclude that there is not a trivial relation between exciton-phonon coupling and the localization of the phonons around the vacancy.
The phonon mode dominating the exciton-phonon coupling spectrum is $\mathrm{\nu=120}$. As shown in Fig. \ref{mode_120}, it is charachterized by a out-of-plane displacement of the atoms around the vacancy, and is peaked in correspondence of the second nearest-neighbours boron atoms. 
\begin{figure}[h]
    \centering
    \includegraphics[scale=0.4]{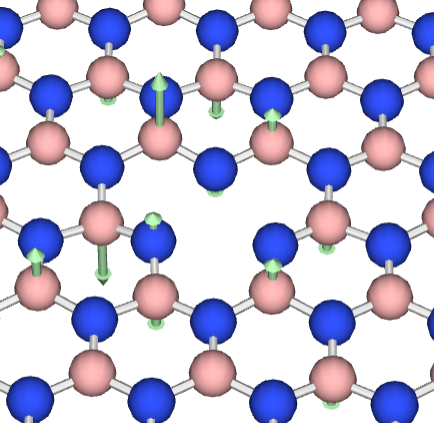}
    \caption{Representation of the eigenvector of the phonon mode $\mathrm{\nu=120}$.}
    \label{mode_120}
\end{figure}

\clearpage
\section*{Dipole selection rules for $D_{3h}$ and $C_{2v}$ groups}
The dipole matrix element between two states $|\psi_i\rangle$ and $|\psi_j\rangle$ is defined as:
\begin{equation}
    \mathbf{d}_{ij} = \langle{\psi_i|\hat{\mathbf{r}}|\psi_j}\rangle
\end{equation}
According to the selection rules from group theory \cite{Tinkham}, the dipole vanishes if the product of the irreducible representations of the wavefunctions $|\psi_{i/j}\rangle$ and the position operator $\hat{\mathbf{r}}$ does not contain the fully symmetric irreducible representation, which is labelled $A'_1$ for the $D_{3h}$ and $A_{1}$ for the $C_{2v}$ one. The product tables for the  $D_{3h}$ and $C_{2v}$ are reported in Tab. \ref{D3h} and \ref{C2v} respectively.\\
For the $D_{3h}$ group, the in plane components (x,y) of $\hat{\mathbf{r}}$ transform as the irreducible representation $E'$, while the out of plane component z as $A''_2$. Tab. \ref{dip_D3h} shows which dipoles matrix elements vanish for symmetry in the symmetric systems.
For the $C_{2v}$ group, the in plane components (x,y) of $\hat{\mathbf{r}}$ transform as the irreducible represenations ($A_1$, $B_1$), while the out of plane component z as $B_2$. Tab. \ref{dip_C2v} shows which dipoles matrix elements vanish for symmetry in the Jahn-Teller distorted systems.

\begin{table}[h]
    \centering
    \begin{tabular}{|c|c|c|c|c|c|c|c|}
    \hline
         & $\mathbf{A'_1}$ & $\mathbf{A'_2}$ & $\mathbf{E'}$ & $\mathbf{A''_1}$ & $\mathbf{A''_2}$ & $\mathbf{E''}$ \\ 
    \hline
     $\mathbf{A'_1}$ & $A'_1$ & $A'_2$ & $E'$ & $A''_1$ & $A''_2$ & $E''$ \\
     \hline
     $\mathbf{A'_2}$ & $A'_2$ & $A'_1$ & $E'$ & $A''_2$ & $A''_1$ & $E''$ \\
     \hline
     $\mathbf{E'}$ & $E'$ & $E'$ & $A'_1+A'_2+E'$ & $E''$ & $E''$ & $A''_1+A''_2+E''$ \\
     \hline
     $\mathbf{A''_1}$ & $A''_1$ & $A''_2$ & $E''$ & $A'_1$ & $A'_2$ & $E'$ \\
     \hline
     $\mathbf{A''_2}$ & $A''_2$ & $A''_1$ & $E''$ & $A'_2$ & $A'_1$ & $E'$ \\
     \hline
     $\mathbf{E''}$ & $E''$ & $E''$ & $A''_1+A''_2 + E''$ & $E'$ & $E'$ & $A'_1 + A'_2 + E'$ \\
     \hline
    \end{tabular}
    \caption{Product table for the $D_{3h}$ group. }
    \label{D3h}
\end{table}
\begin{table}[h]
    \centering
    \begin{tabular}{|c|c|c|c|c|}
    \hline
         & $\mathbf{A_1}$ & $\mathbf{A_2}$ & $\mathbf{B_1}$ & $\mathbf{B_2}$ \\
    \hline
     $\mathbf{A_1}$ & $A_1$ & $A_2$ & $B_1$ & $B_2$ \\
    \hline
     $\mathbf{A_2}$ & $A_2$ & $A_1$ & $B_2$ & $B_1$ \\
    \hline
     $\mathbf{B_1}$ & $B_1$ & $B_2$ & $A_1$ & $A_2$ \\
    \hline
     $\mathbf{B_2}$ & $B_2$ & $B_1$ & $A_2$ & $A_1$ \\
    \hline
    \end{tabular}
    \caption{Product table for the $C_{2v}$ group.}
    \label{C2v}
\end{table}

\begin{table}[]
    \centering
    \begin{tabular}{|c|c|c|c|}
    \hline
        $|\psi_i\rangle$ & $|\psi_j\rangle$ & $d_{xy}$ & $d_z$ \\
    \hline
    \hline
        $A''_2$ & $E'$ & 0 & 0 \\
    \hline
        $E''$ & $E'$ & 0 & $\neq0$ \\
    \hline
        $E''$ & $E'$ & 0 & $\neq0$ \\
    \hline
        $A'_1$ & $E'$ & $\neq0$ & 0 \\
    \hline
        $A''_2$ & $E''$ & $\neq0$ & 0 \\
    \hline 
    \end{tabular}
    \caption{Value of the dipole matrix elements for some transitions of interest in the symmetric system ($D_{3h}$ group).}
    \label{dip_D3h}
\end{table}

\begin{table}[]
    \centering
    \begin{tabular}{|c|c|c|c|}
    \hline
        $|\psi_i\rangle$ & $|\psi_j\rangle$ & $d_{xy}$ & $d_z$ \\
    \hline
    \hline
        $B_2$ & $B_1$ & 0 & 0 \\
    \hline
        $B_2$ & $A_1$ & 0 & $\neq0$ \\
    \hline
    \end{tabular}
    \caption{Value of the dipole matrix elements for some transitions of interest in the Jahn-Teller distorted system ($C_{2v}$ group).}
    \label{dip_C2v}
\end{table}

\clearpage
\section*{Static Jahn-Teller distortion}
\begin{figure}[h]
    \centering
    \begin{subfigure}{0.48\textwidth}
        \includegraphics[scale = 0.5]{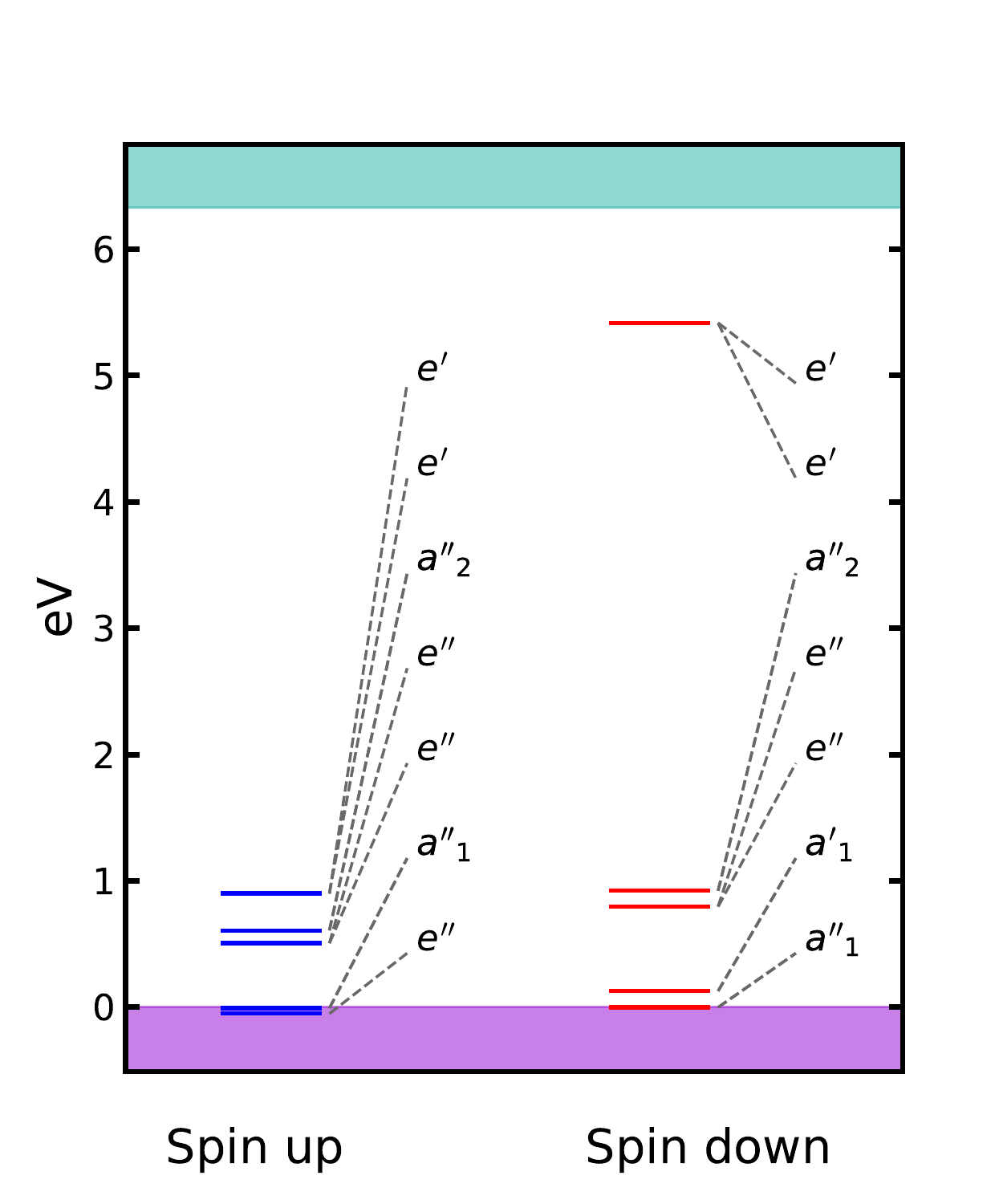}
        \caption{Symmetric system}
        \label{defect_symm}
    \end{subfigure}
   \begin{subfigure}{0.48\textwidth}
            \includegraphics[scale = 0.5]{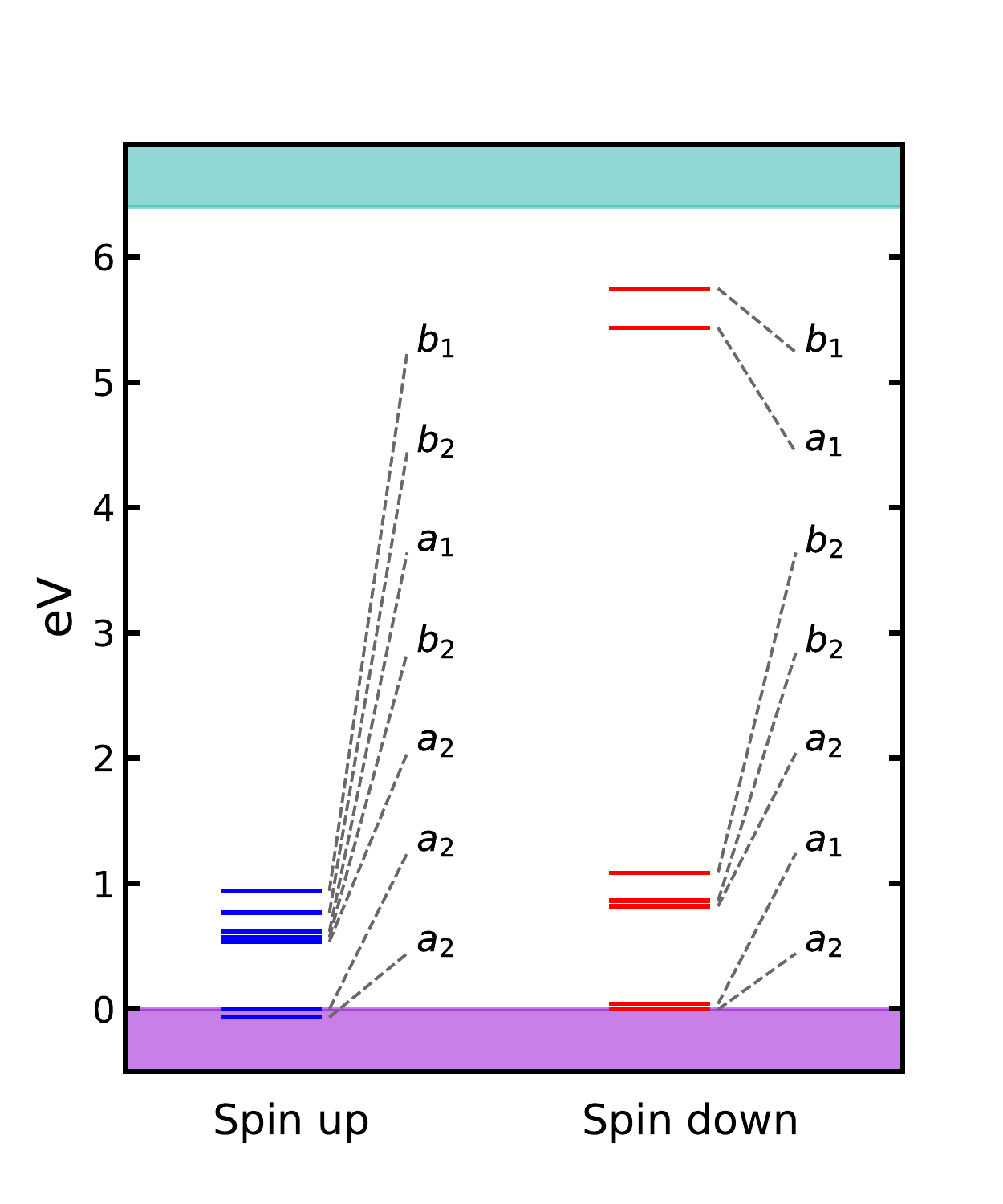}
            \caption{Jahn-Teller distorted}
            \label{defect_asymm}
    \end{subfigure}
    \caption{In panel (a) we replicate Fig. 5.a of the main text, showing the defect levels of the symmetric system calculated at $\mathrm{G_0W_0}$ level of theory. This figure is proposed again here in order to allow a better comparison with the defect levels of the Jahn-Teller distorted system represented in panel (b).}
    \label{defect_levels}
\end{figure}

The doubly degenerate spin down excited state $e'$ can undergo Jahn-Teller distortion, which reduces the symmetry group of the system from $D_{3h}$ to $C_{2v}$. 
As a conquence of it, the doubly degenerate spin down acceptor $e'$ splits into the levels $a_1$ and $b_1$, while the $a''_2$ level is turned into $b_2$ (as shown in Fig. \ref{defect_levels}).
In order to study the Jahn-Teller distorted system, we perform DFT calculations where we relax the system while constraining the occupation of one of the $e'$ orbitals to 1.0 and that of the orbital $a''_2$ to 0. As expected, we find two local minima, corresponding to the two possible configurations in which the doubly degenerate excited state is distorted. One of them has the level $a_1$ lying below the level $b_1$, while in the other this order is switched. In our occupation model, only the lowest lying empty level has a non negligible nonequilibrium population. Therefore, we consider the atomic configuration in which the $a_1$ level is the lowest (Fig. \ref{defect_asymm}), as the transition $a_1\rightarrow b_2$ is allowed by symmetry while the transition $b_1\rightarrow b_2$ is forbidden (see Tab. \ref{dip_C2v}).\\
Since the Jahn-Teller distortion is small, we assume in a first approximation that the exciton energies and wavefunctions remain equal to those of the original symmetric system. The significative difference between the distorted and the symmetric systems is that now the transition $a_1\rightarrow b_2$ ($e'\rightarrow a''_2$) is symmetry-allowed, and this affects the contribution of each exciton to the photoluminescence spectrum. Therefore, in order to check whether the Jahn-Teller effect could be the responsible of the  activation  of the peak at 1.5 eV, we compute the dipole matrix elements corresponding to the transitions $a_1\rightarrow b_2$ and $b_1\rightarrow b_2$ for the distorted system, and use these to recalculate the exciton contributions to photoluminescence.
\\
As commented in the main text, the enhancement of the peak at 1.5 eV due to the static Jahn-Teller effect is very small. It is not difficult to understand the reason: the transition $a_1\rightarrow b_2$ has a non-vanishing out-of-plane dipole that is very small due to the atomically thin material. In addition to this, the Jahn-Teller effect distorts only slightly the geometry, leading to an even weaker dipole moment for the transition.

\clearpage
\section*{Temperature dependence of the phonon contribution to PL}
\begin{figure}[h]
    \centering
    \begin{subfigure}{0.48\textwidth}
            \includegraphics[scale = 0.5]{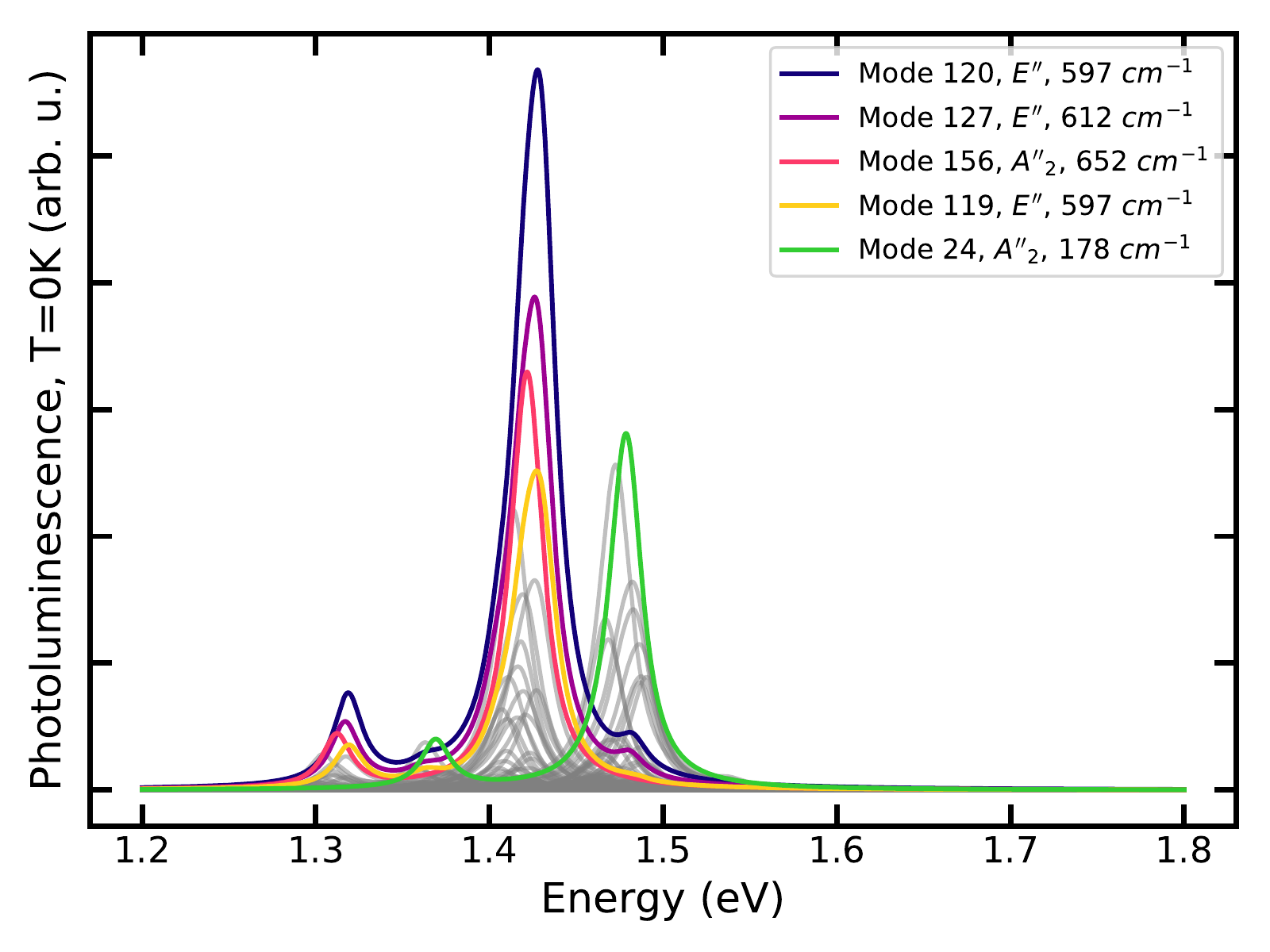}
            \caption{T=0K}
            \label{contribution_77}
    \end{subfigure}
        \begin{subfigure}{0.48\textwidth}
            \includegraphics[scale=0.5]{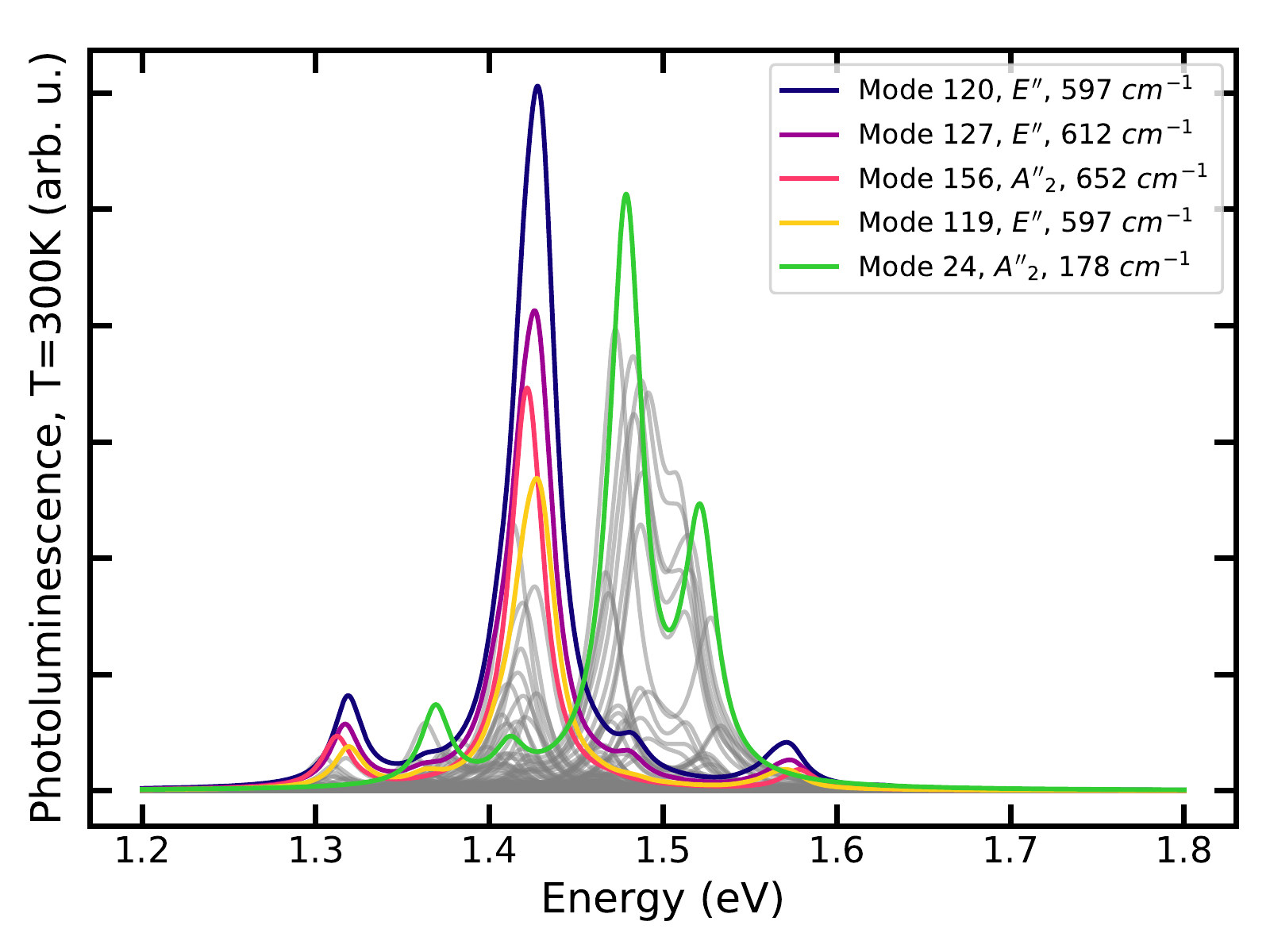}
            \caption{T=300K}
            \label{contribution_300}
    \end{subfigure}
    \caption{Contribution of each phonon to the luminescent emission at T=0K (panel a) and T=300K (panel b). The phonon modes which couple most strongly with excitons are enlightened with coloured lines while the other phonon modes are represented with grey lines.}
    \label{fig:phonon-contribtions}
\end{figure}
If we plot the full contribution of each phonon to the luminescence (as defined in Eq. 2 of the main text), we can see that the factors $\frac{n_B(\omega_{\nu},T)}{2\omega_{\nu}}$ and  $\frac{n_B(\omega_{\nu},T)+1}{2\omega_{\nu}}$, which multiply the  phonon annihilation and the phonon creation peaks respectively, have two main effects. First of all, at low temperatures, the weight corresponding to the phonon creation tends to zero; therefore, only emission at energies lower than $E_{\lambda}$ is expected: Fig. \ref{contribution_77} shows a vanishing contribution to the emission for energies higher than 1.5 eV at T=0K, while it increases remarkably at T=300K. This growth in importance of the phonon creation term with the temperature leads to a broadening of the PL spectrum, as shown in Fig. 3 of the main paper. Furthermore, as the temperature increases, the low-energy phonons tend to contribute more than the high-energy ones. This can be observed by considering the increase in importance of mode $\mathrm{\nu=24}$ of frequency 178 $\mathrm{cm^{-1}}$ (green line), with respect to mode $\mathrm{\nu=120}$ of frequency 570 $\mathrm{cm^{-1}}$ (blue line), leading to the shift towards higher energies of the maximum of the photoluminescence spectrum, as seen in Fig. 3 of the main text.

\clearpage
\section*{Comparison with other experimental spectra}
The theoretical photoluminescence spectrum in Fig. 1 of the main paper has been compared with the experimental curve from Ref. \cite{PhysRevB.102.144105}, which, according to the authors, correspond to isolate emitters. The experimental spectrum of Ref. \cite{Gottscholl2020}, instead, corresponds to an ensemble of spins, and is broader \cite{PhysRevB.102.144105}. It is interesting to compare the theoretical results also with the spectrum of a non-identified defect in 3D hBN reported in Fig. 3b of Ref. \cite{PhysRevApplied.5.034005}, which is characterized by emission at $\mathrm{\sim}$1.6 eV.
\begin{figure}[h]
    \centering
    \includegraphics[scale=0.5]{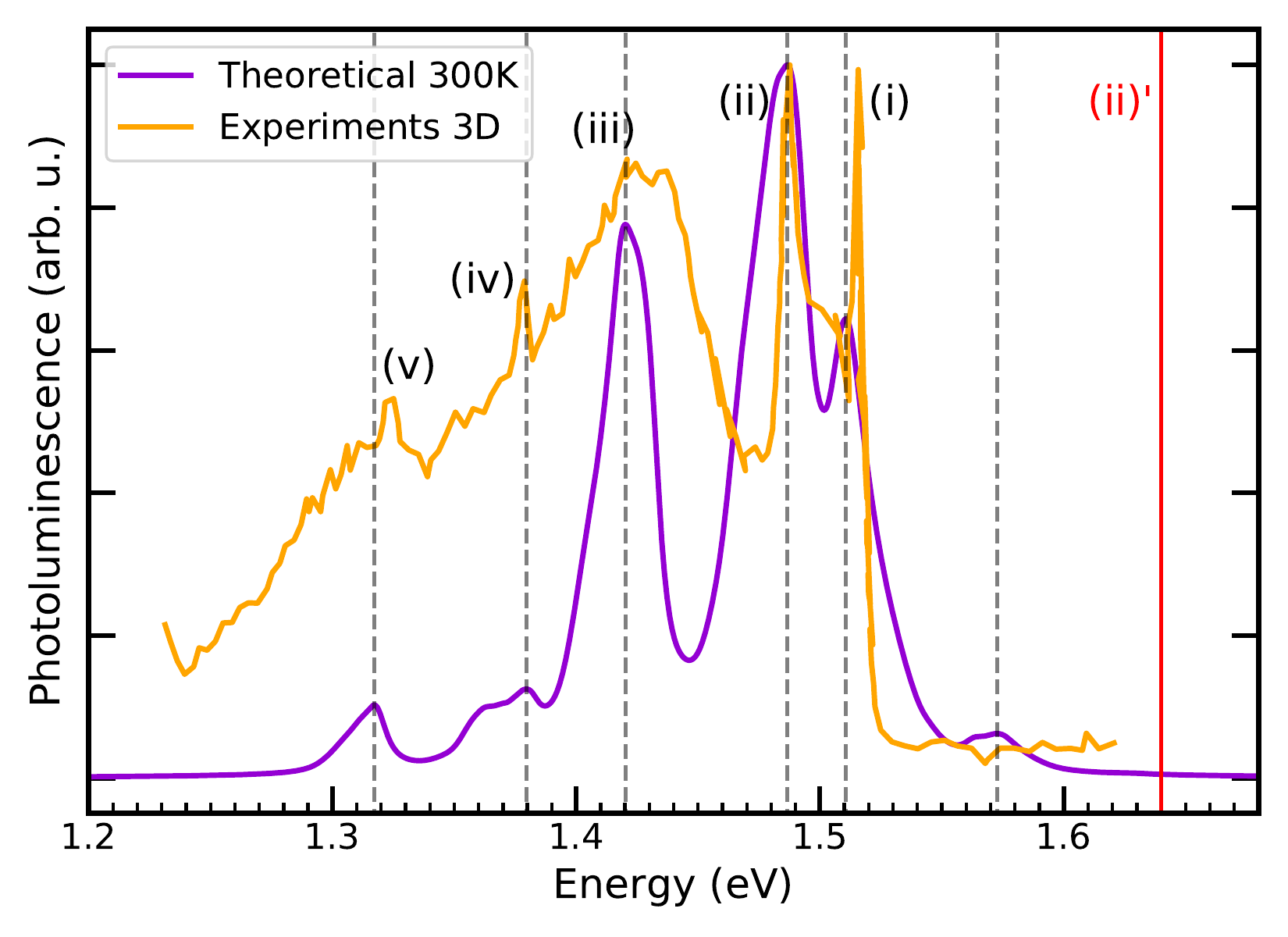}
    \caption{Theoretically predicted photoluminescence spectrum for negatively charged boron vacancy in 2D hBN is compared with the experimental spectrum \cite{PhysRevApplied.5.034005} of a non-identified defect in 3D hBN. The experimental spectrum is aligned with the theoretical one, in such a way that the second peak from the right coincide. The theoretical curve represents the very same spectrum as in Fig. 1 of the main paper, with the only difference that the artificial broadening adopted is very small in this case (0.005 eV, against the 0.04 eV of the figure in the main text) to enlighten the peaks corresponding to the most important phonons. The vertical red line (ii)' corresponds to the energy of the peak (ii) before the shift, and witness a very good agreement also in the absolute position of the theoretical and experimental spectra.}
    \label{bulk}
\end{figure}
Fig. \ref{bulk} reports both this experimental spectrum and the theoretical spectrum for $V_B^-$ in 2D hBN.
According to the authors of Ref. \cite{PhysRevApplied.5.034005}, the peak (i) of this experimental curve is probably due to the background, therefore the highest energy peak corresponding to the defect is the second (ii). The theoretical peak at 1.51 eV corresponds to the annihilation of low frequency phonons, and is weaker than that at 1.49 eV, corresponding to the their creation.     
If we shift the experimental curve in such a way that the most intense peaks coincide (peak (ii) for the experiments and that at 1.49 eV for the theory), we observe a remarkable agreement in the position of the remaining peaks, i.e. (iii), (iv) and (v). In order to realize such alignment, it is necessary to redshift the experimental spectrum of 0.14 eV, witnessing a very good agreement also in the absolute position of the spectra.
Basing on this agreement, we may conclude that:
\begin{enumerate}
    \item the spectrum in Fig. 3b of Ref. \cite{PhysRevApplied.5.034005} is the negatively charged boron vacancy in 3D hBN;
    \item the peak (ii), which was tentatively associated to the ZPL by the authors, correspond instead to the first replica of a phonon of low frequency. 
\end{enumerate}
It must be stressed that an agreement between the spectra corresponding to the same defect in 3D and 2D hBN can be due to the following factors. First of all, the phonons for the 2D hBN look globally similar to those for the 3D hBN, with more marked differences for the LO modes in the long-wavelength limit \cite{Sohier2017}. However, as it can be seen in Fig. \ref{fig:phonon-contribtions}, the contribution of the high frequency phonons to photoluminescence is small.  Furthermore the in-plane dipoles, which are those affecting the most the spectrum, are not likely to change significantly when passing from the 2D to the 3D structure.

\newpage

\section*{Comparison with the Huang-Rhys method}

As anticipated in the main text, the most used technique for the calculation of the PL lineshape of defect centers in semiconductors is based on a generalization of the Huang-Rhys (HR) model \cite{doi:10.1063/1.1700283}. This work assumes that the energy difference between ground and excited state depends on a fictitious coordinate $\xi$ which is linear function of the ground state normal modes. The difference between the relaxed position of the atoms in the excited and in the ground state is thus expanded in term of normal modes to determine the modal coordinates \cite{Alkauskas_2014}
\begin{equation}
    q_k = \sum_{\alpha,i} m_{\alpha}^{1/2}(R_{\alpha i}^e - R_{\alpha i}^g) e_{\alpha i}^k \ ,
\end{equation}
which are later used to calculate the partial Huang-Rhys factor $S_k$ measuring the coupling of the phonon k with the electrons:
\begin{equation} \label{qk}
    S_k = \frac{\omega_k q_k^2}{2\hbar}\ .
\end{equation}
Here $\alpha$ labels the atoms, i refers to the cartesian component, $R^{e/g}_{\alpha i}$ are the equilibrium positions of the atoms in the excited/ground state and  $e_{\alpha i}^k$ is a normalized vector that
describes the displacement of the atom $\alpha$ along the direction i in the phonon mode k, and $\omega_k$ is the phonon frequency.
The function 
\begin{equation}
    S(t) = \int_0^{\infty} \sum_k S_k \delta(\hbar\omega-\hbar\omega_k) e^{-i\omega t} d(\hbar \omega)
\end{equation}
is used to calculate the generating function
\begin{equation}
    G(t)=e^{S(t)-S(0)}
\end{equation}
(see Ref. \cite{2021} for a finite temperature treatment) whose Fourier transform is proportional to the PL lineshape:
\begin{equation}
    L(E_{ZPL}-\hbar \omega) = \omega^3 \int_{-\infty}^{\infty} \frac{1}{2\pi} G(t)e^{i\omega t - \gamma |t|} dt\ .
\end{equation}
Here $\gamma$ is the broadening of the phonon peaks and $E_{ZPL}$ is the zero-phonon line, which is determined through constrained DFT or quantum chemistry calculations. 
We now review the main limitations of this method and see how the many-body perturbation theory (MBPT) treatment of the PL presented in the main text manages to overcome them.
First of all, the starting point of the theory summarized above is the following expression for the absolute intensity of the emission \cite{doi:10.1063/1.1700283,Alkauskas_2014}:
\begin{equation}\label{I}
    I(\hbar \omega) = \frac{n_D \omega^3}{3 \epsilon_0 \pi c^3 \hbar}\  |\mathbf{\mu}_{eg}|^2 \sum_m |\braket{\chi_{gm}|\chi_{e0}}|^2 \ \delta(E_{ZPL}-E_{gm}-\omega\hbar)\ ,
\end{equation}
where $n_D$ is the refractive index, and $|\mathbf{\mu}_{eg}|$ is the optical dipole moment between ground and excited states. This formula is valid for molecules (in which case $n_D=1$) or for solids with homogeneous and isotropic screening, but not for 2d materials, where the screening environment is more complex. In the theory presented in the main text, instead, the whole screening function is calculated through state-of-the-art MBPT calculations \cite{MARINI20091392}. \\ 
The main assumption behind the Huang-Rhys theory is that the optical dipole moment between ground and excited state $|\mathbf{\mu}_{eg}|$ is not affected by phonons (Frank-Condon approximation). This allows to take  $|\mathbf{\mu}_{eg}|$ out of the summation in Eq. \ref{I}. As we will show later this assumption limits strongly the predictivity of the theory. In the many-body perturbation theory, instead phonon dependence of the optical dipole moments is correctly taken into account, as witnessed by the terms $\frac{\partial^2 |\Pi_{\lambda}|^2}{\partial x_{\nu}^2}$ in Eq. 2 of the main text.\\
As shown by Eq. \ref{qk}, only the phonon modes with non-vanishing modal coordinate $q_k$ contributes to the PL lineshape of the Huang-Rhys method. Now, both Refs. \cite{Ivady2020, PhysRevB.102.144105} agree that the observed luminescence is associated to the transition between the many-body states $(1)^3E''\rightarrow (1)^3A'_2$, which corresponds to the transition between the single particle states $e'\rightarrow a''_2$ of the main text. However, the relaxed geometry of the state $(1)^3E''$ remains flat even when considering the Jahn-Teller distortion (i.e. all the atoms have 0 displacement in the out-of-plane direction). This has the fundamental consequence that $q_k=0$ for all the out-of-plane phonon modes: therefore, these modes do not contribute to the HR luminescence. The formalism presented in the main text is beyond the Franck-Condon approximation, and instead predicts that the out-of-plane phonon modes are those affecting most the photoluminescence. 
\begin{figure}[h]
    \centering
    \includegraphics[scale=0.6]{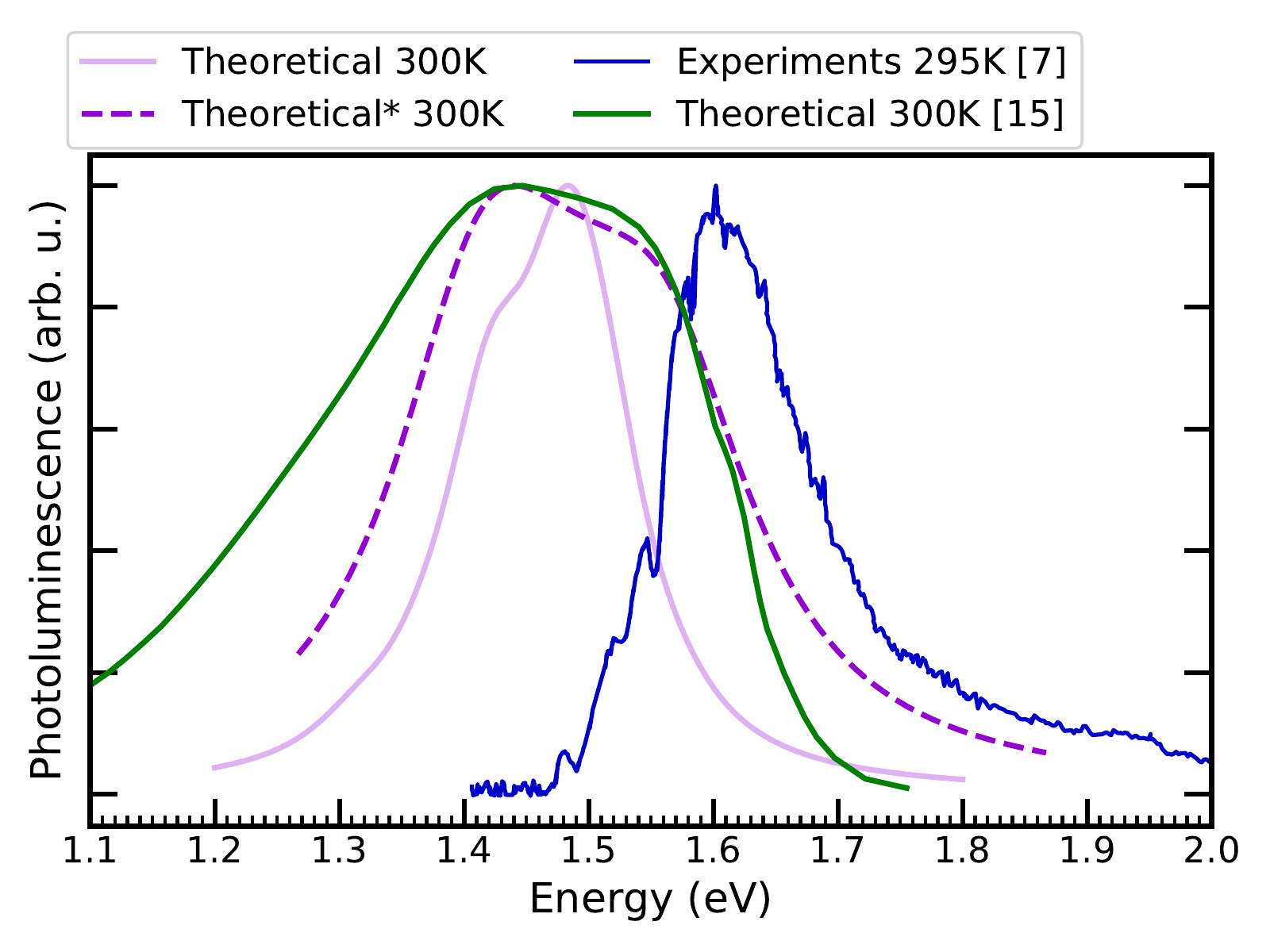}
    \caption{The violet dashed line is obtained from the MBPT approach of the main text when excluding the contribution of the phonon modes of symmetry $A''_1$, $A''_2$ and $E''$. This curve is compared to the result from the HR method of Ref. \cite{Ivady2020}, represented in green solid line. The former curve is shifted in such a way that the main peak coincide with that of the green curve. The solid violet line represents the PL calculated through the MBPT approach when including the contribution of all the phonons, while the solid blue curve represents the experimental PL from Ref. \cite{PhysRevB.102.144105}.  }
    \label{HR}
\end{figure}
It is instructive to calculate the PL with the MBPT approach presented in the main text and excluding the contribution of the phonons of symmetry $A''_1$, $A''_2$ and $E''$ (i.e. those phonons which do not contribute to the PL in the Huang-Rhys method). The result is shown in violet dashed line in Fig. \ref{HR}, and compared with the calculations of the PL through the HR method from Ref. \cite{Ivady2020} (represented in green solid line). In order to facilitate the comparison, the main peak of the two spectra has been aligned. The two spectra are perfectly superimposed in the interval of frequencies ranging from 1.6 eV (where the ZPL of the HR calculation lies, as reported in \cite{Ivady2020}) to 1.4 eV (position of the main peak). The MBPT spectrum decays faster than the HR one for small frequencies, as it does not take into account the harmonic replica. In both cases, the spectrum is much broader than the experimental one \cite{PhysRevB.102.144105}, represented in solid blue line. The result is improved dramatically when the contribution of phonons of all the symmetries is considered when calculating the PL with the MBPT approach, leading to the curve represented in solid violet line (and reported in Fig. 1 of the main text). We thus conclude that the PL is dominated by the phonon modes which are odd with respect to the mirroring with the plane of the 2D material. This explains the failure of the HR method in reproducing the PL lineshape of the charged B vacancy.\\
Regarding the limitation of our method, it does not take into account the polaron shift and the presence of multi-phonon replica. The former can be calculated including also the electron-phonon diagrams in the perturbative series for the one particle and two particles Green's function, while the latter can be determined through by expanding the theory presented in Ref. \cite{PhysRevB.99.081109} to higher orders.

\newpage

%